\documentclass[11pt,a4paper]{article}
\usepackage{jheppub}
\pdfoutput=1

\usepackage[utf8]{inputenc}
\usepackage{amssymb}
\usepackage{amsmath}
\usepackage{amsfonts}
\usepackage{graphicx}
\usepackage{xcolor}
\usepackage{xspace}
\usepackage[normalem]{ulem} 
\usepackage{lscape}
\usepackage{ wasysym }
\usepackage{float}
\usepackage{mathrsfs}
\usepackage{slashed}
\usepackage{multirow,rotating}

\usepackage{hyperref}
\hypersetup{
  colorlinks=true,
  allcolors=darkpurple,
  pdfborder={0 0 0},
  linktocpage=true
}

\definecolor{darkpurple}{rgb}{0.5,0,0.5}
\definecolor{cambridgeblue}{rgb}{0.64, 0.76, 0.68}
\definecolor{darkraspberry}{rgb}{0.53, 0.15, 0.34}

\def\gsim{\raise0.3ex\hbox{$\;>$\kern-0.75em\raise-1.1ex\hbox{$\sim\;$}}}
\def\lsim{\raise0.3ex\hbox{$\;<$\kern-0.75em\raise-1.1ex\hbox{$\sim\;$}}}

\newcommand{\ba}[1]{\begin{eqnarray} \label{(#1)}}
\newcommand{\ea}{\end{eqnarray}}

\def\gsim{\raise0.3ex\hbox{$\;>$\kern-0.75em\raise-1.1ex\hbox{$\sim\;$}}}
\def\lsim{\raise0.3ex\hbox{$\;<$\kern-0.75em\raise-1.1ex\hbox{$\sim\;$}}}

\newcommand{\vtext}[1]{\begin{sideways}#1\end{sideways}}

\makeatletter
\g@addto@macro\bfseries{\boldmath}
\newcommand\Label[1]{&\refstepcounter{equation}(\mathrm{\theequation})\ltx@label{#1}&}
\makeatother

\graphicspath{{figs/}}

\allowdisplaybreaks

\preprint{\begin{flushright} FTUV-22-1005.8958 \\ IFIC/22-27 \end{flushright}}	

\title{Long-lived heavy neutral leptons from mesons in effective field theory}

\author[a]{Rebeca Beltr\'an,}
\emailAdd{rebeca.beltran@ific.uv.es}
\affiliation[a]{{\it AHEP Group, Instituto de F\'{\i}sica Corpuscular --	CSIC/Universitat de Val{\`e}ncia, Apartado 22085,
	E--46071 Val{\`e}ncia, Spain}}

\author[b,c]{Giovanna Cottin,}
\emailAdd{giovanna.cottin@uai.cl}
\affiliation[b]{Departamento de Ciencias, Facultad de Artes Liberales, 
	Universidad Adolfo Ib\'a\~{n}ez,
	Diagonal Las Torres 2640, Santiago, Chile}
\affiliation[c]{Millennium Institute for Subatomic Physics at the High Energy Frontier (SAPHIR), Fernández Concha 700, Santiago, Chile}

\author[d,c]{Juan Carlos Helo,}
\emailAdd{jchelo@userena.cl}
\affiliation[d]{Departamento de F\' isica, Facultad de Ciencias, Universidad de La Serena, 
	Avenida Cisternas 1200, La Serena, Chile }

\author[a]{Martin Hirsch,}
\emailAdd{mahirsch@ific.uv.es}

\author[e]{Arsenii Titov,}
\emailAdd{arsenii.titov@ific.uv.es}
\affiliation[e]{{\it Departament de F{\'i}sica Te{\`o}rica, Universitat de Val{\`e}ncia and Instituto de F\'{\i}sica Corpuscular -- CSIC/Universitat de Val{\`e}ncia, Dr.~Moliner 50, E--46100 Burjassot, Spain}}

\author[f,g]{Zeren~Simon~Wang}
\emailAdd{wzs@mx.nthu.edu.tw}
\affiliation[f]{Department of Physics, National Tsing Hua University, Hsinchu 300, Taiwan}
\affiliation[g]{Center for Theory and Computation, National Tsing Hua University, Hsinchu 300, Taiwan}

\abstract{In the framework of the low-energy effective field theory of
  the Standard Model extended with heavy neutral leptons (HNLs), we
  calculate the production rates of HNLs from meson decays triggered
  by dimension-six operators.  We consider both
  lepton-number-conserving and lepton-number-violating four-fermion
  operators involving either a pair of HNLs or a single HNL.  Assuming
  that HNLs are long-lived, we perform simulations and investigate the
  reach of the proposed far detectors at the high-luminosity LHC to
  (i) active-heavy neutrino mixing and (ii) the Wilson coefficients
  associated with the effective operators, for HNL masses below the
  mass of the $B$-meson.  We further convert the latter to the
  associated new-physics scales.  Our results show that scales
  in excess of hundreds of TeV and the active-heavy mixing squared as small as
  $10^{-15}$ can be probed by these experiments. }

\begin{document}
\maketitle

%



\section{Introduction}
\label{sec:intro}

The next 10--15 years will see a notable increase in experimental
sensitivity to the search for heavy neutral leptons (HNLs), mostly,
but not exclusively, thanks to the diverse long-lived particle
program that has been initiated at CERN for the future
high-luminosity LHC (HL-LHC)~\cite{Alimena:2019zri,Curtin:2018mvb}. Many new
and dedicated ``far-detector'' proposals have been put forward 
in this context: ANUBIS~\cite{Bauer:2019vqk}, AL3X~\cite{Gligorov:2018vkc},
CODEX-b~\cite{Gligorov:2017nwh}, FACET~\cite{Cerci:2021nlb}, FASER and
FASER2~\cite{Feng:2017uoz,FASER:2018eoc}, MoEDAL-MAPP1 and
MAPP2~\cite{Pinfold:2019nqj,Pinfold:2019zwp}, and
MATHUSLA~\cite{Chou:2016lxi,Curtin:2018mvb,MATHUSLA:2020uve}. 
Some of them have already been approved and even deployed. 
In particular, FASER and MoEDAL-MAPP1 will be taking data 
during recently started Run 3 of the LHC, 
whereas the rest of the program is envisioned for the HL phase.

Consequently, a number of papers have studied the discovery prospects
of these (and other) future experiments for minimal HNLs, 
\textit{i.e.} HNLs coupled to the Standard Model~(SM) only through 
active-heavy mixing, see
Refs.~\cite{Kling:2018wct,Helo:2018qej,Dercks:2018wum,Hirsch:2020klk,
  DeVries:2020jbs,Ovchynnikov:2022its}. 
Already in the minimal scenario one can expect up
to four orders of magnitude improvement in mixing angles squared 
relative to existing constraints, for HNLs with masses up to roughly 5 GeV. 
Similarly, also
the main detectors at the LHC can search for HNLs. 
However, ATLAS/CMS will be more
sensitive at larger HNL masses, say, roughly 3--30 GeV; see \textit{e.g.} 
Refs.~\cite{Cottin:2018kmq,Cottin:2018nms,Cheung:2020buy,Jones-Perez:2019plk,
  Liu:2019ayx}. Note that ATLAS~\cite{Aad:2019kiz} as well as CMS~\cite{CMS:2022fut}
have already published first results of searches for HNLs using displaced vertices.

HNLs could appear also in SM extensions with additional
particles, a prominent example being a $Z'$. For sensitivity estimates for this scenario, see \textit{e.g.}~Refs.~\cite{Chiang:2019ajm,Deppisch:2019kvs}. However,
since the LHC has so far not found any unambiguous signal of beyond
the Standard Model (BSM) physics, attention has shifted recently
towards effective field theories (EFTs). 
At the center-of-mass energy of the LHC, one
expects that the correct description is the Standard Model effective field
theory (SMEFT) (see Ref.~\cite{Brivio:2017vri} for a review) 
or, if we allow also for the existence of light (with masses below or around the electroweak scale) 
right-handed (RH) gauge singlet fermions $N_R$, 
the $N_R$SMEFT~\cite{delAguila:2008ir,Aparici:2009fh,Liao:2016qyd,Li:2021tsq}. 
Recently, we have studied the expectation for
HNL searches in the $N_R$SMEFT for pair-$N_R$ operators~\cite{Cottin:2021lzz} and for single-$N_R$ operators~\cite{Beltran:2021hpq}.

However, below the electroweak scale, and in particular, at the scale of meson masses, 
the massive gauge and Higgs bosons, as well as the top quark, 
have to be integrated out. The correct EFT 
in this lower energy regime is then usually called LEFT (low-energy effective field theory)~\cite{Jenkins:2017jig}, or in case we add RH singlet states, 
$N_R$LEFT~\cite{Bischer:2019ttk,Chala:2020vqp,Li:2020lba,Li:2020wxi}.
The sensitivities of future far detectors at the LHC for 
the four-fermion $N_R$LEFT operators containing a single $N_R$, 
a charged lepton $\ell$, and a quark bilinear, and thus, triggering decays 
of mesons (produced in $pp$ collisions at the LHC) into $N_R$ and $\ell$,
have already been studied in Ref.~\cite{DeVries:2020jbs}. 
Furthermore, the projected sensitivities of Belle II to the same set of operators 
have been derived in Ref.~\cite{Zhou:2021ylt} (see also Ref.~\cite{Han:2022uho}).
Therefore, in the current work, we will concentrate on pair-$N_R$ operators and operators with
one $N_R$ and one SM neutrino, 
exploring the potential of future far detectors at the LHC.
We note that Ref.~\cite{Li:2020lba} has studied a similar set of operators and derived bounds from current data on the 
coherent elastic neutrino-nucleus scattering (CE${\nu}$NS) process 
and invisible decays of light unflavored mesons.

The rest of this paper is organized as follows. In the next section, we
discuss $N_R$LEFT operators at $d=6$ and their matching to
the $N_R$SMEFT at $d=6$ and $d=7$. 
Then in section~\ref{sec:hnlproduction}, 
we elaborate on HNL production from meson decays 
in the framework of the $N_R$LEFT. 
Section~\ref{sec:simulation} gives a brief summary of the experiments we consider and provides the corresponding simulation details. 
In section~\ref{sec:results}, we present the results of our simulation. 
We then close with a brief summary. 
Technical details of the computations of meson and HNL 
decay widths in the $N_R$LEFT are relegated to the appendices.

\section{Effective field theory approach}
\label{sec:eft}
%
%
\subsection{$N_R$LEFT: low-energy effective field theory with right-handed neutrinos}
\label{sec:NRLEFT}
%
In the present work, we are interested in the HNL production from meson decays 
in an EFT approach. 
Meson decays take place at energy scales $E \sim \mathcal{O}(1)$~GeV, 
much smaller than the electroweak scale $v = 246$~GeV. 
At such energies, the $W$ and $Z$ gauge bosons, the Higgs boson and the top quark are not propagating degrees of freedom, 
and the effects of potential new physics (NP) associated with a scale $\Lambda \gg E$ 
can be described in terms of the LEFT extended with RH neutrinos (or HNLs), $N_R$. 
We will refer to such minimal extension of the LEFT as $N_R$LEFT. 
This EFT respects $SU(3)_C \times U(1)_\mathrm{em}$ gauge symmetry 
and in addition to the gluons and the photon contains all SM leptons, 
$u$, $d$, $s$, $c$, $b$ quarks and, in general, an arbitrary number $n_N$ of HNLs.

In addition to the renormalizable terms and higher-dimensional operators 
constructed from the SM fields that have been systematically classified up to $d=6$ in Ref.~\cite{Jenkins:2017jig}, the $N_R$LEFT Lagrangian includes operators with $N_R$, which at $d \leq 6$ have been constructed in Refs.~\cite{Chala:2020vqp,Li:2020lba}. 
The renormalizable part of the Lagrangian is given by
\begin{equation}
 \mathcal{L}_\mathrm{ren} = \mathcal{L}_\mathrm{QCD+QED} 
 + \overline{N_R} i \slashed{\partial} N_R 
 - \left[\frac{1}{2} \overline{\nu_L} M_\nu \nu_L^c 
 + \frac{1}{2} \overline{N_R^c} M_N N_R 
 + \overline{\nu_L} M_D N_R + \text{h.c.}\right],
\end{equation}
where $\mathcal{L}_\mathrm{QCD+QED}$ is the QCD and QED Lagrangian 
built out of the SM leptons (including left-handed (LH) neutrinos $\nu_L$ without any mass term) and the five aforementioned flavors of quarks 
(for its form, see \textit{e.g.}~Eq.~(5.1) in Ref.~\cite{Jenkins:2017jig}). 
The superscript $c$ denotes a charge conjugate field, 
$\psi^c \equiv C \overline{\psi}^T$. 
The Majorana mass matrices $M_\nu$ and $M_N$ are complex symmetric
$3 \times 3$ and $n_N \times n_N$ matrices, respectively. 
$M_D$ is a generic $3 \times n_N$ Dirac mass matrix. 
If lepton number is conserved, the Majorana mass terms are not present.
The full $N_R$LEFT Lagrangian reads
\begin{equation}
 \mathcal{L}_{N_R\mathrm{LEFT}} = \mathcal{L}_\mathrm{ren} 
 + \sum_{d \geq 5} \sum_{i} c_i^{(d)} \mathcal{O}_i^{(d)}\,,
\end{equation}
where $c_i^{(d)}$ are the Wilson coefficients of higher-dimensional operators $\mathcal{O}_i^{(d)}$, and the second sum goes over 
all independent operators of a given mass dimension $d$. 
The dimensionful coefficients $c_i^{(d)}$ scale as $\Lambda^{4-d}$. 

At $d=5$, there are two dipole operators with $N_R$:
\begin{equation}
 \mathcal{O}_{NN\gamma} = \overline{N_R^c} \sigma^{\mu\nu} N_R F_{\mu\nu} 
 \qquad
 \text{and}
 \qquad
 \mathcal{O}_{\nu N \gamma} = \overline{\nu_L} \sigma^{\mu\nu} N_R F_{\mu\nu}\,,
\end{equation}
where $F_{\mu\nu}$ is the photon field strength tensor. 
The first of these operators is antisymmetric in the HNL generation indices, 
and thus, it vanishes identically for $n_N = 1$. 
$\mathcal{O}_{NN\gamma}$ is lepton-number-violating (LNV), 
whereas $\mathcal{O}_{\nu N \gamma}$ is lepton-number-conserving (LNC). 
The phenomenology of the RH neutrino dipole operator 
has been studied in Ref.~\cite{Aparici:2009fh} and more recently 
in Ref.~\cite{Barducci:2022gdv} in the context of the proposed 
far detectors at the LHC.

At $d=6$, there are 55 four-fermion interactions, 
of which 23 LNC~\cite{Chala:2020vqp}, 
26 LNV, and six operators that violate both lepton and baryon numbers~\cite{Li:2020lba}.%
\footnote{Here, we do not count possible flavor structures and Hermitian conjugates.}
The contact operators triggering meson decays into HNLs involve a pair of quarks and a leptonic part, which in turn contains either
(i) a pair of HNLs, or (ii) $N_R$ and $\nu_L$, or else (iii) $N_R$ and a charged lepton. The last possibility has been investigated in detail in Refs.~\cite{DeVries:2020jbs} and \cite{Zhou:2021ylt}, 
in the context of long-lived HNLs that could be produced 
in meson decays at the high-luminosity LHC and Belle~II, respectively. 
In this work, we focus on the operators with neutral leptons. 
In tables~\ref{tab:opsNN} and \ref{tab:opsN}, we summarize 
the pair-$N_R$ and single-$N_R$ operators of interest. 
\begin{table}[t]  
\renewcommand{\arraystretch}{1.3}
\centering
 \begin{tabular}[t]{|c|c|c|c|c|}
    \cline{2-5}
    \multicolumn{1}{c|}{} & Name & Structure& $n_N = 1$ & $n_N = 3$ \\ 
    \cline{2-5}
    \noalign{\vskip\doublerulesep\vskip-\arrayrulewidth}
    \hline
    \multirow{4}{*}{\vtext{LNC}} & ${\cal O}_{dN}^{V,RR}$ &
    $\left(\overline{d_R}\gamma_{\mu}d_R\right)\left(\overline{N_R}\gamma^{\mu}N_R\right)$ & 
     9 &
     81 \\
    & ${\cal O}_{uN}^{V,RR}$ &
    $\left(\overline{u_R}\gamma_{\mu}u_R\right)\left(\overline{N_R}\gamma^{\mu}N_R\right)$ &
    4 & 
    36 \\
    \cline{2-5}
    & ${\cal O}_{dN}^{V,LR}$ &
    $\left(\overline{d_L}\gamma_{\mu}d_L\right)\left(\overline{N_R}\gamma^{\mu}N_R\right)$ & 
    9 &
    81 \\
    & ${\cal O}_{uN}^{V,LR}$ &
    $\left(\overline{u_L}\gamma_{\mu}u_L\right)\left(\overline{N_R}\gamma^{\mu}N_R\right)$ & 
    4 &
    36 \\
     \hline
 %
 %
    \hline
    \multirow{6}{*}{\vtext{LNV}} & ${\cal O}_{dN}^{S,RR}$ &
    $\left(\overline{d_L}d_R\right)\left(\overline{N_R^c}N_R\right)$ & 
    18 &
    108 \\
    & ${\cal O}_{dN}^{T,RR}$ &
    $\left(\overline{d_L}\sigma_{\mu\nu}d_R\right)\left(\overline{N_R^c}\sigma^{\mu\nu}N_R\right)$ & 
     0 &
     54 \\
    & ${\cal O}_{uN}^{S,RR}$ &
    $\left(\overline{u_L}u_R\right)\left(\overline{N_R^c}N_R\right)$ &
    8 & 
    48 \\
    & ${\cal O}_{uN}^{T,RR}$ &
    $\left(\overline{u_L}\sigma_{\mu\nu}u_R\right)\left(\overline{N_R^c}\sigma^{\mu\nu}N_R\right)$ &
    0 & 
    24 \\
    \cline{2-5}
    & ${\cal O}_{dN}^{S,LR}$ &
    $\left(\overline{d_R}d_L\right)\left(\overline{N_R^c}N_R\right)$ & 
     18 &
     108 \\
     & ${\cal O}_{uN}^{S,LR}$ &
    $\left(\overline{u_R}u_L\right)\left(\overline{N_R^c}N_R\right)$ & 
     8 &
     48 \\
     \hline
 \end{tabular}
 \caption{Dimension-six operators in the $N_R$LEFT, involving two quarks and two HNLs. For each operator structure, we give the number of
independent real parameters for $n_N = 1$ and $n_N = 3$ 
 generations of $N_R$. We recall that in the $N_R$LEFT, $n_d = 3$ and $n_u = 2$. 
 The LNV operator structures require ``+h.c.''.}
 \label{tab:opsNN}
\end{table}
%
%
%
%
\begin{table}[t]  
\renewcommand{\arraystretch}{1.3}
\centering
 \begin{tabular}[t]{|c|c|c|c|c|}
    \cline{2-5}
    \multicolumn{1}{c|}{} & Name & Structure & $n_N = 1$ & $n_N = 3$ \\ 
    \cline{2-5}
    \noalign{\vskip\doublerulesep\vskip-\arrayrulewidth}
    \hline
    \multirow{6}{*}{\vtext{LNC}} & ${\cal O}_{d\nu N}^{S,RR}$ &
    $\left(\overline{d_L}d_R\right)\left(\overline{\nu_L}N_R\right)$ & 54
      & 162
      \\
    & ${\cal O}_{d\nu N}^{T,RR}$ &
    $\left(\overline{d_L}\sigma_{\mu\nu}d_R\right)\left(\overline{\nu_L}\sigma^{\mu\nu}N_R\right)$ &  54
      & 162
      \\
    & ${\cal O}_{u\nu N}^{S,RR}$ &
    $\left(\overline{u_L}u_R\right)\left(\overline{\nu_L}N_R\right)$ & 24
     & 72
     \\
    & ${\cal O}_{u\nu N}^{T,RR}$ &
    $\left(\overline{u_L}\sigma_{\mu\nu}u_R\right)\left(\overline{\nu_L}\sigma^{\mu\nu}N_R\right)$ &  24
     & 72
     \\
     \cline{2-5}
     & ${\cal O}_{d\nu N}^{S,LR}$ &
    $\left(\overline{d_R}d_L\right)\left(\overline{\nu_L}N_R\right)$ &  54
    & 162
     \\
     & ${\cal O}_{u\nu N}^{S,LR}$ &
    $\left(\overline{u_R}u_L\right)\left(\overline{\nu_L}N_R\right)$ & 24
    & 72
     \\
     \hline
 %
 %
    \hline
    \multirow{4}{*}{\vtext{LNV}} & ${\cal O}_{d\nu N}^{V,RR}$ &
    $\left(\overline{d_R}\gamma_{\mu}d_R\right)\left(\overline{\nu_L^c}\gamma^{\mu}N_R\right)$ &  54
     & 162
     \\
    & ${\cal O}_{u\nu N}^{V,RR}$ &
    $\left(\overline{u_R}\gamma_{\mu}u_R\right)\left(\overline{\nu_L^c}\gamma^{\mu}N_R\right)$ & 24
      & 72
      \\
    \cline{2-5}
    & ${\cal O}_{d\nu N}^{V,LR}$ &
    $\left(\overline{d_L}\gamma_{\mu}d_L\right)\left(\overline{\nu_L^c}\gamma^{\mu}N_R\right)$ & 54
      & 162
      \\
     & ${\cal O}_{u\nu N}^{V,LR}$ &
    $\left(\overline{u_L}\gamma_{\mu}u_L\right)\left(\overline{\nu_L^c}\gamma^{\mu}N_R\right)$ & 24
      & 72
      \\
     \hline
 \end{tabular}
 \caption{Dimension-six operators in the $N_R$LEFT, involving two quarks, one active neutrino and one HNL. For each operator structure, we give the number of
independent real parameters for $n_N = 1$ and $n_N = 3$ 
 generations of $N_R$. 
 We recall that in the $N_R$LEFT, $n_\nu = n_d = 3$ and $n_u = 2$. 
 All operator structures require ``+h.c.''.}
 \label{tab:opsN}
\end{table}
%
Though in our numerical analysis we will focus on the case of a single HNL generation ($n_N = 1$), in the tables we provide for completeness the numbers of independent real parameters associated with each operator for $n_N = 1$ and $n_N = 3$. 
We have checked these numbers using the \texttt{Sym2Int} package~\cite{Fonseca:2017lem,Fonseca:2019yya}. 
In the group of pair-$N_R$ operators, there are 
four LNC structures and six LNV ones (of which two vanish identically for $n_N = 1$). 
Among the single-$N_R$ operators, six structures conserve lepton number, 
whereas four violate it.

\subsection{Matching to the $N_R$SMEFT}
\label{sec:matching}
%
In this subsection, we assume that the $N_R$LEFT originates from the $N_R$SMEFT,%
\footnote{It is worth noting that ($N_R$)LEFT is the correct effective description 
at low energies even if the high-energy EFT is not given by the ($N_R$)SMEFT but by the HEFT (with $N_R$), 
in which the Higgs is not part of a fundamental weak doublet~\cite{Jenkins:2017jig}.}
 \textit{i.e.}~the EFT of the SM extended with RH neutrinos~\cite{delAguila:2008ir,Aparici:2009fh,Liao:2016qyd}. 
The $N_R$SMEFT respects the SM gauge symmetry 
$SU(3)_C \times SU(2)_L \times U(1)_Y$, and it is valid at energies 
above the electroweak scale, set by the vacuum expectation value (VEV)
of the Higgs, $v$. 
In the considered case of the $N_R$LEFT as the low-energy limit of the $N_R$SMEFT, there are no new particles at energies between $E$ and $v$, and the $N_R$LEFT scale $\Lambda$ can be identified with $v$.
At this scale, the two EFTs must be matched.

At tree level, the matching of the $d=6$ LNC operators 
has been worked out in Ref.~\cite{Chala:2020vqp} and that of the $d=6$ LNV operators in Ref.~\cite{Li:2020lba}.%
\footnote{For the $N_R$LEFT operators with two quarks, 
a charged lepton and a neutrino, the tree-level matching has been performed also in Refs.~\cite{Dekens:2020ttz,DeVries:2020jbs}. 
For the first-generation charged fermions, such operators contribute, 
among other processes, to neutrinoless double beta decay~\cite{Dekens:2020ttz,Dekens:2021qch,deVries:2022nyh,Cirigliano:2022oqy}.}
Below we summarize the results of this matching relevant to the operators of interest. 
The $N_R$SMEFT operators contributing to the matching relations 
for the coefficients of the $N_R$LEFT operators from tables~\ref{tab:opsNN} and \ref{tab:opsN}
are given in tables~\ref{tab:NSMEFTopsNN} and \ref{tab:NSMEFTopsN}, respectively.
\begin{table}[t] 
\renewcommand{\arraystretch}{1.3}
\centering 
 \begin{tabular}{|c|c|c|c|c|}
    \cline{2-5}
    \multicolumn{1}{c|}{} & Name & Structure & $n_N = 1$ & $n_N = 3$ \\ 
    \cline{2-5}
    \noalign{\vskip\doublerulesep\vskip-\arrayrulewidth}
    \hline
    \multirow{4}{*}{\vtext{$d=6$ (LNC)}} & ${\cal O}_{dN}$ &
    $\left(\overline{d_R}\gamma_{\mu}d_R\right)\left(\overline{N_R}\gamma^{\mu}N_R\right)$ & 
     9 &
     81 \\
    & ${\cal O}_{uN}$ &
    $\left(\overline{u_R}\gamma_{\mu}u_R\right)\left(\overline{N_R}\gamma^{\mu}N_R\right)$ &
    9 & 
    81 \\
    & ${\cal O}_{QN}$ &
    $\left(\overline{Q}\gamma_{\mu}Q\right)\left(\overline{N_R}\gamma^{\mu}N_R\right)$ & 
    9 &
    81 \\
     & ${\cal O}_{HN}$ &
    $\left(H^\dagger i \overleftrightarrow{D_\mu} H\right)\left(\overline{N_R} \gamma^\mu N_R\right)$ & 
    1 &
    9 \\
     \hline
 %
 %
    \hline
    \multirow{4}{*}{\vtext{$d=7$ (LNV)}} & ${\cal O}_{QNdH}$ &
    $\left(\overline{Q}N_R\right)\left(\overline{N_R^c}d_R\right)H$ & 18
     & 162
     \\
   & ${\cal O}_{dQNH}$ &
    $H^\dagger \left(\overline{d_R}Q\right)\left(\overline{N_R^c}N_R\right)$ & 18
     & 108
      \\
    & ${\cal O}_{QNuH}$ &
    $\left(\overline{Q}N_R\right)\left(\overline{N_R^c}u_R\right)\widetilde{H}$ & 18    & 162
    \\
    & ${\cal O}_{uQNH}$ &
    $\widetilde{H}^\dagger \left(\overline{u_R}Q\right)\left(\overline{N_R^c}N_R\right)$ & 18
     & 108
     \\
    \hline
 \end{tabular}
 \caption{Dimension-six and dimension-seven operators in the $N_R$SMEFT contributing to the tree-level matching relations for the Wilson coefficients 
 of the dimension-six $N_R$LEFT operators from table~\ref{tab:opsNN}. 
 For each operator structure, we give the number of
independent real parameters for $n_N = 1$ and $n_N = 3$ 
 generations of $N_R$. The LNV operator structures require ``+h.c.''.}
 \label{tab:NSMEFTopsNN}
\end{table}
%
%
%
%
\begin{table}[t] 
\renewcommand{\arraystretch}{1.3}
\centering 
 \begin{tabular}{|c|c|c|c|c|}
    \cline{2-5}
    \multicolumn{1}{c|}{} & Name & Structure & $n_N = 1$ & $n_N = 3$ \\ 
    \cline{2-5}
    \noalign{\vskip\doublerulesep\vskip-\arrayrulewidth}
    \hline
    \multirow{3}{*}{\vtext{$d=6$ (LNC)}} & ${\cal O}_{LNQd}$ &
 $\epsilon_{ab} \left(\overline{L^a}N_R\right) \left(\overline{Q^b}d_R\right)$ &
  54 & 
  162 \\
  & ${\cal O}_{LdQN}$ & 
  $\epsilon_{ab} \left(\overline{L^a}d_R\right) \left(\overline{Q^b}N_R\right)$ &
   54 & 
   162 \\
  &${\cal O}_{QuNL}$ &
  $\left(\overline{Q}u_R\right)\left(\overline{N_R}L\right)$ &
  54 & 
  162 \\
     \hline
 %
 %
    \hline
    \multirow{6}{*}{\vtext{$d=7$ (LNV)}} & ${\cal O}_{dNLH}$ &
     $\epsilon_{ab} \left(\overline{d_R} \gamma_\mu d_R\right) \left(\overline{N_R^c} \gamma^\mu L^a\right) H^b$ & 
     54 & 
     162 \\
   & ${\cal O}_{uNLH}$  &
     $\epsilon_{ab} \left(\overline{u_R} \gamma_\mu u_R\right) \left(\overline{N_R^c} \gamma^\mu L^a\right) H^b$ & 
     54 & 
    162 \\
    & ${\cal O}_{QNLH1}$ &
    $\epsilon_{ab} \left(\overline{Q} \gamma_\mu Q\right) \left(\overline{N_R^c} \gamma^\mu L^a\right) H^b$ & 
    54 & 
    162 \\
    & ${\cal O}_{QNLH2}$ &
    $\epsilon_{ab} \left(\overline{Q} \gamma_\mu Q^a\right) \left(\overline{N_R^c} \gamma^\mu L^b\right) H$ & 
     54 & 
     162 \\
   & ${\cal O}_{NL1}$ &
   $\epsilon_{ab} \left(\overline{N_R^c} \gamma_\mu L^a\right) \left(iD^\mu H^b\right) \left(H^\dagger H\right)$ &
   6 &
   18 \\  
   & ${\cal O}_{NL2}$ &
   $\epsilon_{ab} \left(\overline{N_R^c} \gamma_\mu L^a\right) H^b \left(H^\dagger i \overleftrightarrow{D^\mu} H\right)$ &
   6 &
   18 \\  
   \hline
 \end{tabular}
 \caption{Dimension-six and dimension-seven operators in the $N_R$SMEFT contributing to the tree-level matching relations for the Wilson coefficients 
 of the dimension-six $N_R$LEFT operators from table~\ref{tab:opsN}. 
 For each operator structure, we give the number of
independent real parameters for $n_N = 1$ and $n_N = 3$ 
 generations of $N_R$. All operator structures require ``+h.c.''.}
 \label{tab:NSMEFTopsN}
\end{table}
For the dimensionful Wilson coefficients of the $N_R$SMEFT operators, 
we use capital $C_i^{(d)}$. They scale as $\Lambda_\mathrm{NP}^{4-d}$, 
where $\Lambda_\mathrm{NP} \gg v$ is the scale of new physics. 
For simplicity, we assume only one generation of $N_R$, 
while considering a general structure of quark, charged lepton and LH neutrino flavor indices. 
We denote the quark flavor indices by $i\,,j$ and the lepton flavor index by $\alpha$. 
Furthermore, we assume that in the $N_R$LEFT, quarks are in the mass basis, whereas the $N_R$SMEFT operators are written in the weak interaction basis.

The matching conditions for the coefficients of the LNC pair-$N_R$ operators 
holding at the electroweak scale read:
\begin{align}
 c_{dN,ij}^{V,RR} &= C_{dN}^{ij} - \frac{g_Z^2}{m_Z^2} Z_{d_R}^{ij} Z_N\,,  \hspace{1.75cm} 
 c_{uN,ij}^{V,RR} = C_{uN}^{ij} - \frac{g_Z^2}{m_Z^2} Z_{u_R}^{ij} Z_N\,, \\
 c_{dN,ij}^{V,LR} &= V_{ki}^\ast V_{lj} C_{QN}^{kl} - \frac{g_Z^2}{m_Z^2} Z_{d_L}^{ij} Z_N\,, \qquad
 c_{uN,ij}^{V,LR} = C_{QN}^{ij} - \frac{g_Z^2}{m_Z^2} Z_{u_L}^{ij} Z_N\,.
\end{align}
Here, $V$ is the CKM matrix defined through $d'_L = V d_L$, 
where $d'$ and $d$ are flavor and mass eigenstates, respectively.
In addition, 
\begin{equation}
 g_Z \equiv \frac{e}{s_W c_W}\,,
 \qquad
 Z_\psi^{ij} \equiv \left(T_\psi^3 - Q_\psi s_W^2\right) \delta^{ij}\,, 
 \qquad 
 Z_N \equiv - \frac{v^2}{2} C_{HN}\,,
\end{equation}
with $\psi$ denoting a SM fermion, and $T_\psi^3$ and $Q_\psi$ being
its weak isospin and electric charge, respectively. 
Further,  
$s_W \equiv \sin\theta_W$ and $c_W \equiv \cos\theta_W$, 
where $\theta_W$ is the weak mixing angle, 
and $m_Z$ is the mass of the $Z$ boson.
Note that in $Z_\psi^{ij}$ we neglect the contributions of $d=6$
$N_R$SMEFT operators (cf.~Eq.~(2.18) in Ref.~\cite{Li:2020lba}), 
since when multiplied by $Z_N$ these would lead to $d=8$ effects.

For the LNC single-$N_R$ operators, we have:
\begin{align}
 c_{d\nu N, i j \alpha}^{S,RR} &= V_{ki}^\ast \left(C_{LNQd}^{\alpha k j} - \frac{1}{2} C_{LdQN}^{\alpha j k}\right), \qquad
 c_{d\nu N, i j \alpha}^{T,RR} = - \frac{1}{8} V_{ki}^\ast C_{LdQN}^{\alpha j k}\,, \\
 c_{u\nu N, i j \alpha}^{S,RR} &= 0\,, \hspace{4.8cm}
 c_{u\nu N, i j \alpha}^{T,RR} = 0\,, \\
 c_{d\nu N, i j \alpha}^{S,LR} &= 0\,, \hspace{4.8cm}
 c_{u\nu N, i j \alpha}^{S,LR} = C_{QuNL}^{j i \alpha\, \ast}\,.
\end{align}
We note that the Higgs exchange contributions (which a priori seem relevant) 
are parametrically of the same order 
as $d=8$ terms, and thus, can be dropped when working to $d=6$~\cite{Jenkins:2017jig}.

In the $N_R$SMEFT at $d=6$, there is only one operator which violates lepton number, 
but conserves baryon number, namely, $\mathcal{O}_{NNNN} = (\overline{N_R^c} N_R) (\overline{N_R^c} N_R)$. It vanishes identically in the case of one generation of $N_R$. 
Thus, the $d=6$ LNV operators in the $N_R$LEFT are expected to get contributions 
from $d=7$ LNV operators in the $N_R$SMEFT. 
A basis of the latter can be found in Ref.~\cite{Liao:2016qyd} 
(see also Ref.~\cite{Bhattacharya:2015vja}). 
A subset of the $d=7$ operators relevant for the matching we are interested in
is collected in the bottom part of tables~\ref{tab:NSMEFTopsNN} and \ref{tab:NSMEFTopsN}. 

For the LNV pair-$N_R$ operators, the matching conditions read:
\begin{align}
 c_{dN,ij}^{S,RR} &= -\frac{v}{2\sqrt2} V_{ki}^\ast C_{QNdH}^{kj}\,, \qquad 
 c_{uN,ij}^{S,RR} = -\frac{v}{2\sqrt2} C_{QNuH}^{ij}\,, 
 \label{eq:matching_LNV_pairN} \\ 
 c_{dN,ij}^{S,LR} &= \frac{v}{\sqrt2} V_{kj} C_{dQNH}^{ik}\,, \hspace{1.3cm}
 c_{uN,ij}^{S,LR} = \frac{v}{\sqrt2} C_{uQNH}^{ij}\,.
\end{align}

Finally, for the LNV single-$N_R$ operators, we have:
\begin{align}
 c_{d\nu N, i j \alpha}^{V,RR} &= -\frac{v}{\sqrt2} C_{dNLH}^{ij\alpha} - \frac{g_Z^2}{m_Z^2} Z_{d_R}^{ij} Z_{\nu N}^\alpha\,, \\
 c_{u\nu N, i j \alpha}^{V,RR} &= -\frac{v}{\sqrt2} C_{uNLH}^{ij\alpha} - \frac{g_Z^2}{m_Z^2} Z_{u_R}^{ij} Z_{\nu N}^\alpha\,, \\
 c_{d\nu N, i j \alpha}^{V,LR} &= -\frac{v}{\sqrt2} V_{ki}^\ast V_{lj}
 \left(C_{QNLH1}^{kl\alpha} - C_{QNLH2}^{kl\alpha}\right) - \frac{g_Z^2}{m_Z^2} Z_{d_L}^{ij} Z_{\nu N}^\alpha\,, \\
 c_{u\nu N, i j \alpha}^{V,LR} &= -\frac{v}{\sqrt2} C_{QNLH1}^{ij\alpha} - \frac{g_Z^2}{m_Z^2} Z_{u_L}^{ij} Z_{\nu N}^\alpha\,.
\end{align}
Here, the couplings $Z_{\nu N}^\alpha$ are defined as
\begin{equation}
 Z_{\nu N}^\alpha \equiv \frac{v^3}{4\sqrt2} \left(C_{NL1}^\alpha + 2 C_{NL2}^\alpha\right).
\end{equation}
%

\subsection{Running of the $N_R$LEFT operators of interest}
\label{sec:running}
%
Between the energy scale at which meson decays take place and the weak scale, 
the Wilson coefficients of the $N_R$LEFT operators evolve according to 
renormalization group equations (RGEs). 
These are analogous to the RGEs in the LEFT (without $N_R$)~\cite{Jenkins:2017dyc}, 
and some partial results for the operators with $N_R$ can be found in 
Refs.~\cite{Chala:2020vqp,DeVries:2020jbs,Li:2020lba}. 
Below we provide the one-loop RGEs for the pair-$N_R$ and single-$N_R$ operators 
summarized in tables~\ref{tab:opsNN} and~\ref{tab:opsN}. 
The operators given by the product of vector currents 
are renormalized by QED corrections:%
\footnote{In these equations, we ignore the contributions from the four-fermion leptonic operators $\mathcal{O}_{eN}^{V,RR}$ and $\mathcal{O}_{eN}^{V,LR}$ 
as well as from the double insertions of the $d=5$ neutrino dipole operators.}
\begin{align}
 \dot{c}_{qN,ij}^{V,RR} &=  \dot{c}_{qN,ij}^{V,LR} = \frac{4}{3} e^2 Q_q  N_c \delta_{ij} 
 \left[Q_u \left(c_{uN,kk}^{V,RR} + c_{uN,kk}^{V,LR}\right) + 
 Q_d \left(c_{dN,kk}^{V,RR} + c_{dN,kk}^{V,LR}\right)\right], \\
 \dot{c}_{q\nu N,ij\alpha}^{V,RR} &= \dot{c}_{q\nu N,ij\alpha}^{V,LR} 
 = \frac{4}{3} e^2 Q_q  N_c \delta_{ij} 
 \left[Q_u \left(c_{u\nu N,kk\alpha}^{V,RR} + c_{u\nu N,kk\alpha}^{V,LR}\right) + 
 Q_d \left(c_{d\nu N,kk\alpha}^{V,RR} + c_{d\nu N,kk\alpha}^{V,LR}\right)\right]. 
\end{align}
Here, $\dot{c} \equiv 16 \pi^2 \mu \dfrac{\mathrm{d}c}{\mathrm{d}\mu}$, with $\mu$ being the renormalization scale; 
 $q = d$ or $u$; $e$ is the QED coupling constant, $Q_q$ is the electric charge 
of the quark $q$; and $N_c = 3$ is the number of colors. 
The flavor index $k$ runs over $u$ and $c$ for the up-type quarks, 
and over $d,s,b$ for the down-type quarks. 
We see that only the coefficients with $i=j$ run, whereas those with $i\neq j$, 
which will be of interest to us in what follows, do not.

The operators of the scalar and tensor types are renormalized 
due to both QCD and QED corrections:
\begin{align}
 \dot{c}_q^S &= -6\left(g^2 C_F + e^2 Q_q^2\right) c_q^S\,, 
 \qquad
 c_q^S = c_{qN,ij}^{S,RR}\,,~c_{qN,ij}^{S,LR}\,,~c_{q\nu N,ij\alpha}^{S,RR}\,,~c_{q\nu N,ij\alpha}^{S,LR}\,, \\
 \dot{c}_q^T &= 2\left(g^2 C_F + e^2 Q_q^2\right) c_q^T\,, 
 \qquad
 c_q^T = c_{qN,ij}^{T,RR}\,,~c_{q\nu N,ij\alpha}^{T,RR}\,,
\end{align}
where $g$ is the strong coupling constant, and $C_F = (N_c^2-1)/(2N_c) = 4/3$. 

Solving these RGEs to the leading-logarithm approximation, we find: 
\begin{align}
 c_q^S(\mu) &= \left[\frac{g(\mu)}{g(\mu_1)}\right]^\frac{6 C_F}{b_g} 
 \left[\frac{e(\mu)}{e(\mu_1)}\right]^\frac{6 Q_q^2}{b_e} c_q^S(\mu_1)\,, \\
 c_q^T(\mu) &= \left[\frac{g(\mu)}{g(\mu_1)}\right]^{-\frac{2 C_F}{b_g}} 
 \left[\frac{e(\mu)}{e(\mu_1)}\right]^{-\frac{2 Q_q^2}{b_e}} c_q^T(\mu_1)\,,
\end{align}
where $\mu_1 < \mu$ are two energy scales, 
and $b_g$ and $b_e$ are the coefficients of the one-loop QCD and QED beta functions, respectively:%
\footnote{For the beta functions, we use $\dot{g} = - b_g g^3$ and $\dot{e} = - b_e e^3$.}
\begin{align}
 b_g &= \frac{11}{3}N_c - \frac{2}{3}\left(n_u+n_d\right), \\
 b_e &= -\frac{4}{3} \left(Q_e^2 n_e + Q_d^2 N_c n_d +Q_u^2 N_c n_u\right).
\end{align}
Setting $\mu_1 = m_b$ and $\mu = v$, we find 
\begin{align}
 c_u^S(v) &= 0.68\, c_u^S(m_b)\,, \qquad c_d^S(v) = 0.69\, c_d^S(m_b)\,, \\
 c_u^T(v) &= 1.14\, c_u^T(m_b)\,, \qquad c_d^T(v) = 1.13\, c_d^T(m_b)\,.
\end{align}
Thus, when running from $m_b$ to $v$, the scalar type operators 
get suppressed, whereas the tensor type operator get enhanced, 
cf. Ref.~\cite{Li:2020lba}. 
Finally, we note that the contribution to the running from QED corrections 
is practically negligible.

Concerning the running in the $N_R$SMEFT,
\textit{i.e.} between the EW scale and the scale of new physics, 
$\Lambda_\mathrm{NP}$, the one-loop RGEs for $d=6$ operators with $N_R$ 
have been derived in Refs.~\cite{Chala:2020pbn,Datta:2020ocb,Datta:2021akg}, while partial results for some $d=7$ $N_R$-operators (relevant for neutrinoless double beta decay) have been obtained in Ref.~\cite{Dekens:2020ttz}.

\subsection{An example UV completion for $d=7$ $N_R$SMEFT}
\label{sec:uv}
%
The operators including one or two $N_R$ in the $N_R$SMEFT can be generated 
in the ultra-violet (UV) in a variety of ways. For example, Ref.~\cite{Bischer:2019ttk} discusses all possible leptoquark states that can induce $d=6$ $N_R$SMEFT operators. Also Refs.~\cite{Cottin:2021lzz} 
and \cite{Beltran:2021hpq} study some example models for pair-$N_R$ 
and single-$N_R$ operators at $d=6$, respectively, including leptoquark 
and $Z'$ models. However, neither of these references gives example 
models for the $d=7$ $N_R$SMEFT operators, to which our low-energy 
LNV $N_R$ operators must be matched. 

Here, we do not attempt to give a systematic list of such models for
the $d=7$ operators.  Instead, we will briefly discuss one concrete,
but very typical UV model as an example. This example 
model generates the operator ${\cal O}_{QNdH}$. For the other $d=7$ operators 
containing one or two $N_R$, other UV models can be constructed in an 
analogous manner.

Our example model contains two scalar leptoquark states, 
$S_1 \sim (\mathbf{3},\mathbf{1})_{-1/3}$ and $S_2 \sim (\mathbf{3},\mathbf{2})_{1/6}$, with the numbers inside parentheses denoting 
the  $(SU(3)_C,SU(2)_L)$ representations and the subscript standing for the hypercharge.
Apart from the mass and kinetic terms for the 
leptoquark states, the model allows to write down the following 
interactions, in addition to the SM Lagrangian: 
\begin{equation}
 \mathcal{L} \supset - \left[ Y_{QN} \overline{Q}N_R S_2 
 + Y_{Nd} \overline{N_R^c}d_R S_1^{\dagger} 
 + \mu H S_ 2^{\dagger} S_1 + {\rm h.c.} \right] + \cdots,
  \label{eq:lagUV}
\end{equation}
where, for simplicity, we have suppressed generation indices. Note 
that this Lagrangian violates lepton number, but lepton number 
is recovered in the limit $\mu\rightarrow 0$ \cite{Hirsch:1996qy}.
Integrating out both leptoquarks will produce ${\cal O}_{QNdH}$ with 
$C_{QNdH} = ( \mu Y_{QN}Y_{Nd})/(m_{S_1}^2m_{S_2}^2)$. In the limit 
$\mu = m_{S_1}=m_{S_2}=\Lambda_\mathrm{NP}$, this gives the correct 
$C_{QNdH} \propto \Lambda_\mathrm{NP}^{-3}$ dependence for a $d=7$ operator, 
as expected. Clearly, this model will also generate $d=6$ operators 
with $N_R$. For example, the term proportional to $Y_{QN}$ ($Y_{Nd}$) times 
its hermitian conjugate will generate ${\cal O}_{QN}$ (${\cal O}_{dN}$). 
However, these are only examples, as actually more $d=6$ operators can be 
generated in this model; note the dots in Eq.~\eqref{eq:lagUV}.


\section{HNL production in meson decays}
\label{sec:hnlproduction}
%
We are interested in HNLs produced in meson decays. 
In the framework of the $N_R$LEFT, such decays are triggered 
by both active-heavy neutrino mixing and higher-dimensional operators. 
The HNL production via mixing has been studied in detail in the literature, 
see \textit{e.g.} Refs.~\cite{Bondarenko:2018ptm,Ballett:2019bgd,Coloma:2020lgy} 
for recent reappraisals. 
The HNL production via effective interactions 
has been considered in Ref.~\cite{DeVries:2020jbs} 
for the four-fermion operators with two quarks, 
one charged lepton and $N_R$.
In this section, we focus on the meson decay modes triggered by the
operators with two quarks 
and two neutral leptons (either a pair of $N_R$, or $N_R$ and $\nu_L$) 
discussed in section~\ref{sec:eft}.

Given their large production rates at the LHC, we focus on pseudoscalar 
mesons containing either a $c$ or $b$ quark and a light quark, \textit{i.e.} $D$- and $B$-mesons. Vector mesons with the same quark content have 
much smaller lifetimes, and thus, their contribution in the HNL production 
is sub-dominant~\cite{DeVries:2020jbs}.

\begin{figure}[t]
 \centering
 \includegraphics[width=0.48\textwidth]{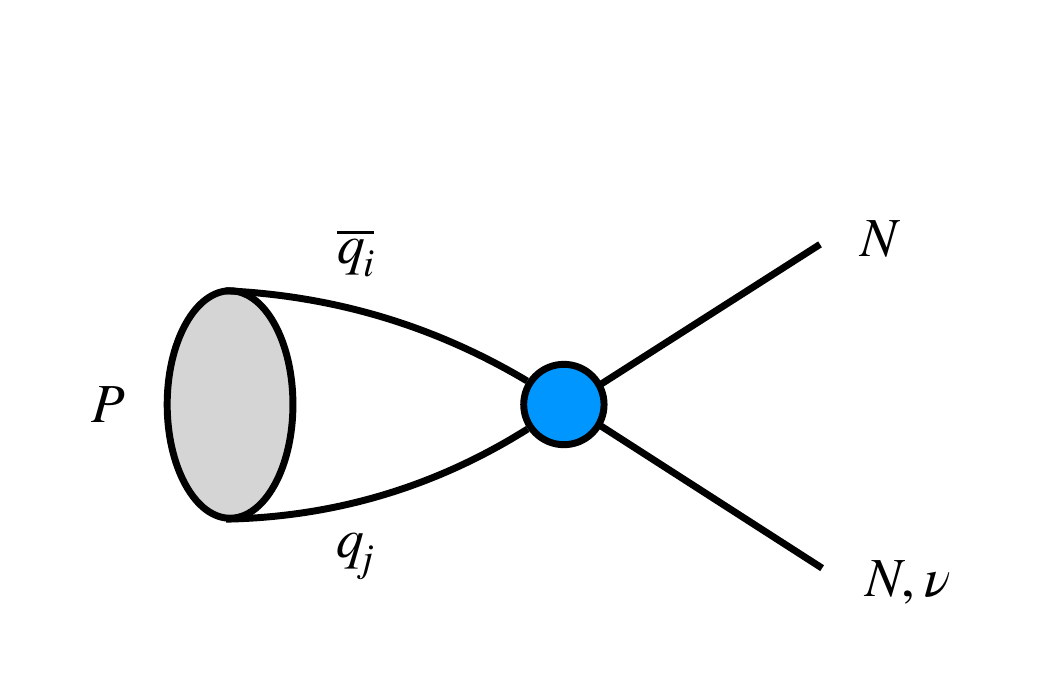}
 \hspace{2mm}
 \includegraphics[width=0.48\textwidth]{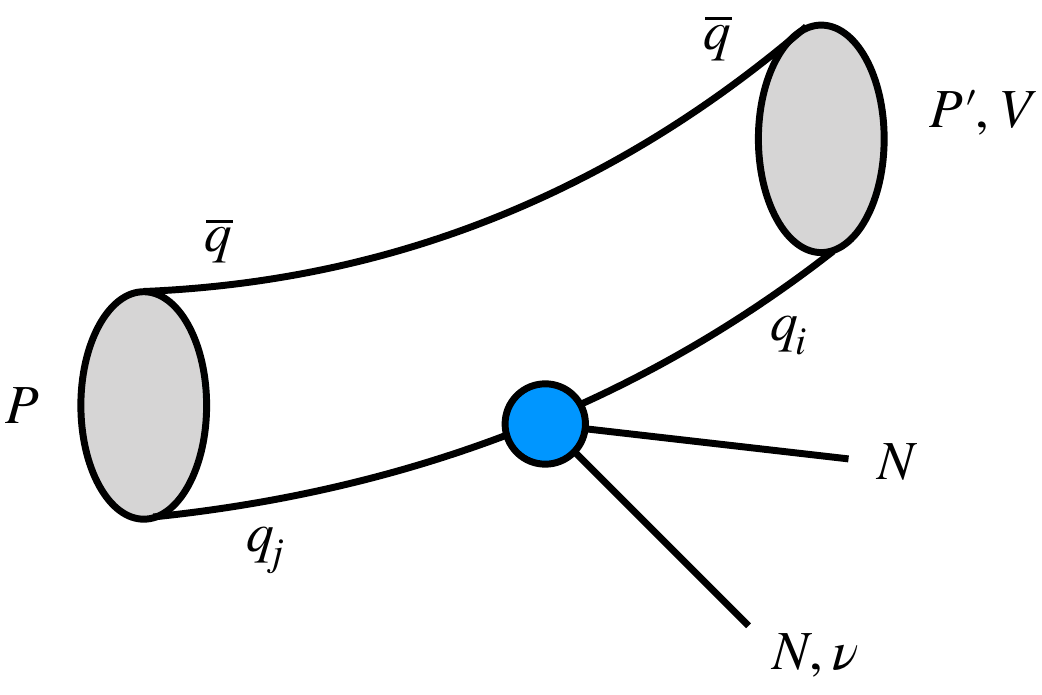}
 \caption{Two- and three-body pseudoscalar meson decays 
 triggered by the effective operators with a pair of HNLs, 
or one HNL and one SM neutrino. 
 The blue blob denotes the insertion of an effective operator. 
 In the right diagram, the quark $q$ can be $u$, $d$, $s$, $c$ or $b$, 
 whereas the quarks $q_i$ and $q_j$ are both of either the up- or down-type, 
 with $m_{q_j} > m_{q_i}$. $P^{(\prime)}$ and $V$ denote pseudoscalar and vector mesons, respectively.}
 \label{fig:decays}
\end{figure}

In figure~\ref{fig:decays}, we depict two- and three-body 
pseudoscalar meson decays triggered by the effective operators 
of interest.
%
The same $N_R$LEFT operator induces a leptonic two-body decay 
as well as contributes to a number of semi-leptonic three-body decays.
In the diagram corresponding to the three-body decay, 
$q$ can be any quark but top, since the latter does not form bound states 
and is not present in the spectrum of the $N_R$LEFT. 
For the transition $q_j \to q_i$ with $i \neq j$, both quarks have to be of 
either the up- or down-type, since the operators considered 
involve quark bilinears with zero electric charge.
In addition, $m_{q_j} > m_{q_i}$ for the decay to be kinematically allowed.%
\footnote{This condition is necessary, but not sufficient, since a meson mass is not a simple sum of the masses of its constituent quarks.} 
These considerations lead to the following four possibilities: 
$c \to u$ (for the up-quark sector), and 
$b \to d$, $b \to s$ and  $s \to d$ (for the down-quark sector). 
Below we discuss each of them in detail.
\begin{itemize}
 \item The operators with $q_j = c$ and $q_i = u$ lead to the leptonic 
 two-body decays $D^0 \to NN$ and $D^0 \to \nu N$.%
 \footnote{In the case of Dirac $\nu$ and $N$, these decays should read 
 $D^0 \to N\overline{N}$ and $D^0 \to \nu \overline{N}$ ($D^0 \to \overline{\nu} N$).}
 In addition, they generate the following transitions resulting in semi-leptonic  three-body decays: 
 $D^0 \to \pi^0\,,~\eta\,,~\eta'~(\rho^0\,,~\omega)$ for $q=u$; 
 $D^+ \to \pi^+~(\rho^+)$ for $q=d$; 
 $D_s^+ \to K^+~(K^{\ast+})$ for $q=s$; 
 $\eta_c \to \overline{D^0}~(\overline{D^{\ast0}})$ for $q=c$; and
 $B_c^+ \to B^+~(B^{\ast+})$ for $q=b$.
 Here, the final states outside (inside) parentheses are pseudoscalar (vector) mesons. In what follows, we will not take into account the decays of $\eta_c$ 
 and $B_c^+$. The $\eta_c$ meson has a lifetime of $\mathcal{O}(10^{-23})$~s~\cite{Workman:2022ynf}, 
 which is much smaller than that of other pseudoscalar mesons 
 and comparable with the lifetimes of vector mesons. 
 Concerning $B_c^+$, its production rate at the LHC is much smaller 
 than that of $D$- and $B$-mesons, 
see \textit{e.g.}~Ref.~\cite{LHCb:2019tea}.
 \item The operators with $q_j = b$ and $q_i = d$~%
 \footnote{In figure~\ref{fig:decays}, we need to send $\overline{q} \to q$ and $q_{j,i} \to \overline{q_{j,i}}$ to match the particle content of $B$-mesons.}
 induce 
 $B^0 \to NN$ and $B^0 \to \nu N$ decays, 
 as well as the following transitions: 
 $B^+ \to \pi^+~(\rho^+)$ for $q=u$; 
 $B^0 \to \pi^0\,,~\eta\,,~\eta'~(\rho^0\,,~\omega)$ for $q=d$; 
 $B_s^0 \to \overline{K^0}~(\overline{K^{\ast 0}})$ for $q=s$;
 $B_c^+ \to D^+~(D^{\ast+})$ for $q=c$; and
 $\eta_b \to \overline{B^0}~(\overline{B^{\ast0}})$ for $q=b$. 
 Again, we will not consider the decays of $B_c^+$ as well as those of $\eta_b$.
 \item The operators with $q_j = b$ and $q_i = s$ give rise to 
 $B_s^0 \to NN$ and $B_s^0 \to \nu N$ decays. They also contribute to
 $B^+ \to K^+~(K^{\ast+})$ for $q=u$;
 $B^0 \to K^0~(K^{\ast0})$ for $q=d$; 
 $B_s^0 \to \eta\,,~\eta'~(\phi)$ for $q=s$;
 $B_c^+ \to D_s^+~(D_s^{\ast+})$ for $q=c$; and
 $\eta_b \to \overline{B_s^0}~(\overline{B_s^{\ast0}})$ for $q=b$.
 \item The operators with $q_j = s$ and $q_i = d$ trigger 
 $K^0 \to NN$ and $K^0 \to \nu N$ decays, 
 as well as the following transitions: 
 $K^+ \to \pi^+
 $ for $q=u$; 
 $K^0 \to \pi^0
 $ for $q=d$;
 $\eta' \to \overline{K^0}~(\overline{K^{\ast0}})$ for $q=s$;
 $D_s^+ \to D^+
 $ for $q=c$; and 
 $\overline{B_s^0} \to \overline{B^0}~(\overline{B^{\ast0}})$ for $q=b$. 
 We will not consider these decay modes in what follows. 
This is because for the listed $D_s^+$ and $\overline{B_s^0}$ decays, the available mass window is very small, and for the light mesons' decays (kaons and $\eta'$), the numerical predictions from Pythia8~\cite{Sjostrand:2014zea,Bierlich:2022pfr}, which we will use for numerical simulation in this work, do not agree well with the LHCf experiment~\cite{pythiaForward}.
\end{itemize}

We have computed the corresponding partial 
decay widths assuming both $N$ and $\nu$ 
are (i) Dirac and (ii) Majorana particles. 
In appendix~\ref{sec:decay_formulae}, 
we provide the formulae for the two-body decay widths 
and the three-body decay amplitudes in both cases. 
To compute the three-body decay widths, we have used 
the procedure described in Ref.~\cite{DeVries:2020jbs}. 
We summarize it in appendix~\ref{app:3-body} for completeness.

If the products of HNL decays are not detected, 
the modes we consider here 
contribute to the $P \to~\text{inv.}$ or $P \to P'/V \nu \overline{\nu}$ decays. 
The existing upper limits on the branching ratios of such decays are listed in table~\ref{tab:invisible}.
In what follows, we will take these constraints into account.
\begin{table}[t]
\renewcommand{\arraystretch}{1.2}
\centering
 \begin{tabular}{| l | c | l | c | l | c |}
 \hline
 Decay & Limit on BR & Decay & Limit on BR & Decay & Limit on BR \\
 \hline
 \hline
 $D^0 \to \text{inv.}$ & \textbf{$9.4 \times 10^{-5}$} & 
 $B^0 \to \text{inv.}$ & \textbf{$2.4 \times 10^{-5}$} & 
 $B_s^0 \to \phi \nu \overline{\nu}$ & $5.4 \times 10^{-3}$ \\
  & & 
 $B^0 \to \pi^0 \nu \overline{\nu}$ & \textbf{$9.0 \times 10^{-6}$} & 
 $B^0 \to K^0 \nu \overline{\nu}$ & $2.6 \times 10^{-5}$ \\
  & & 
 $B^0 \to \rho^0 \nu \overline{\nu}$ & $4.0 \times 10^{-5}$ & 
 $B^0 \to K^{\ast 0} \nu \overline{\nu}$ & $1.8 \times 10^{-5}$ \\
 & & 
$B^+ \to \pi^+ \nu \overline{\nu}$ & $1.4 \times 10^{-5}$ & 
$B^+ \to K^+ \nu \overline{\nu}$ & \textbf{$1.6 \times 10^{-5}$} \\
  & & 
 $B^+ \to \rho^+ \nu \overline{\nu}$ & $3.0 \times 10^{-5}$ & 
 $B^+ \to K^{\ast +} \nu \overline{\nu}$ & $4.0 \times 10^{-5}$ \\
 \hline
 \end{tabular}
 \caption{Upper limits on the branching ratios (BRs) of (semi-)invisible $D$- and $B$-meson decays extracted from Ref.~\cite{Workman:2022ynf}. The limits highlighted in bold provide the most stringent constraints on the corresponding Wilson coefficients, see section~\ref{sec:results}.}
 \label{tab:invisible}
\end{table}
%

\begin{figure}[t]
 \centering
 \includegraphics[height=0.29\textheight]{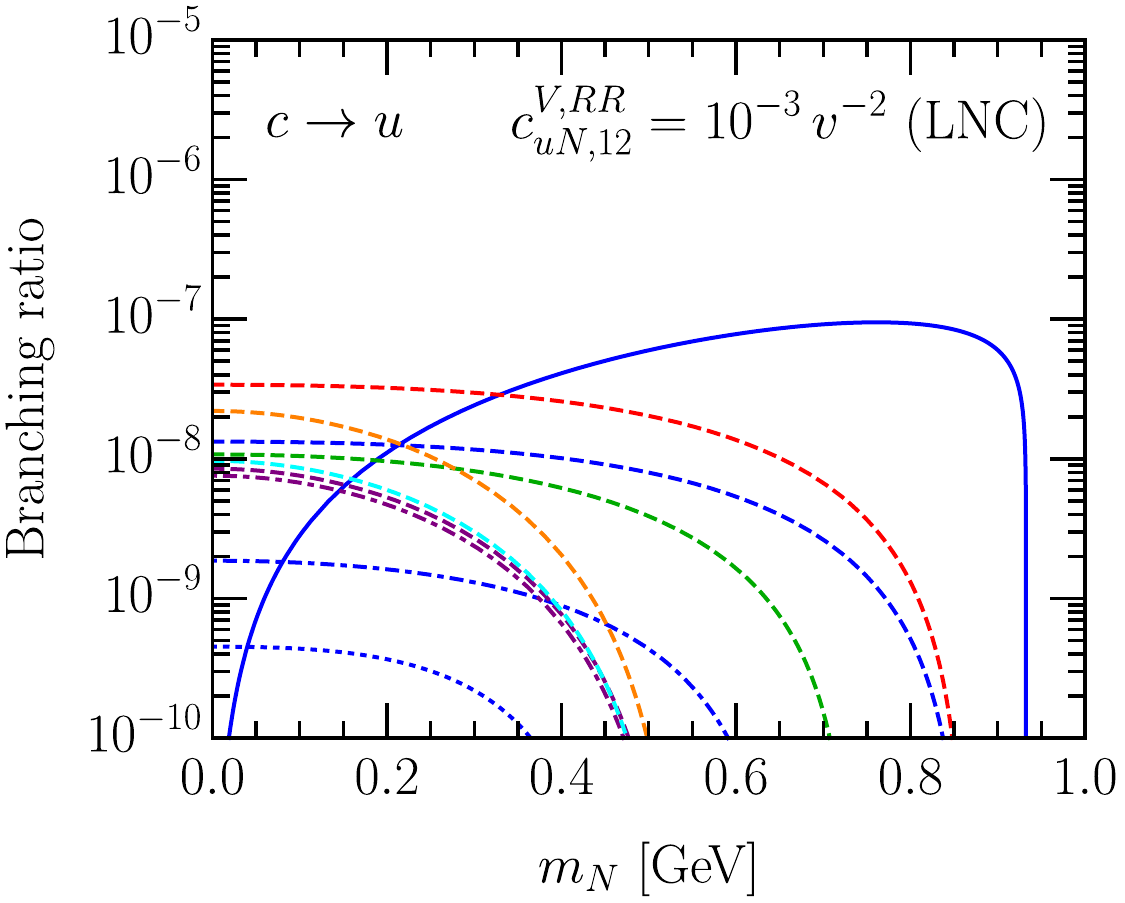}
 \includegraphics[height=0.29\textheight]{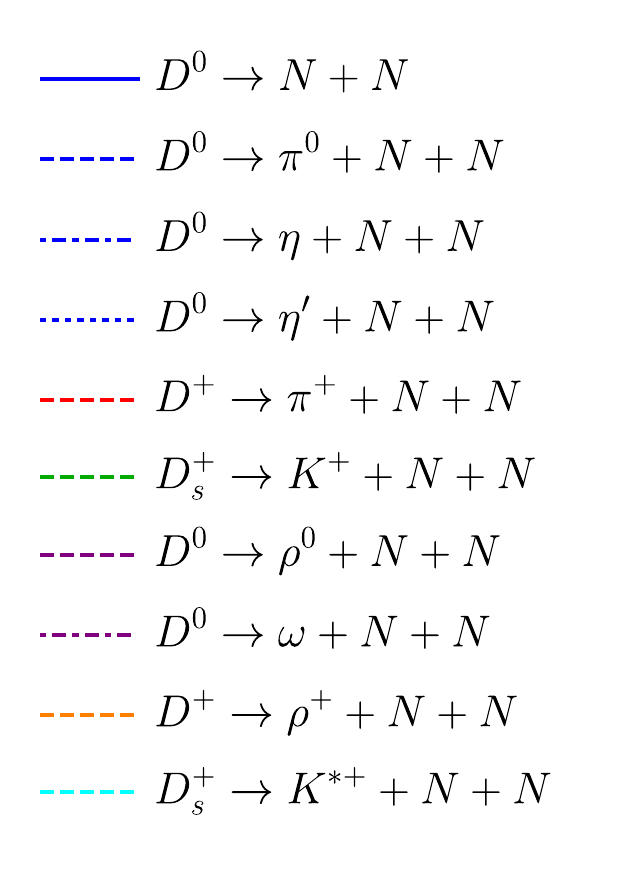}\\[0.15cm]
 \includegraphics[height=0.29\textheight]{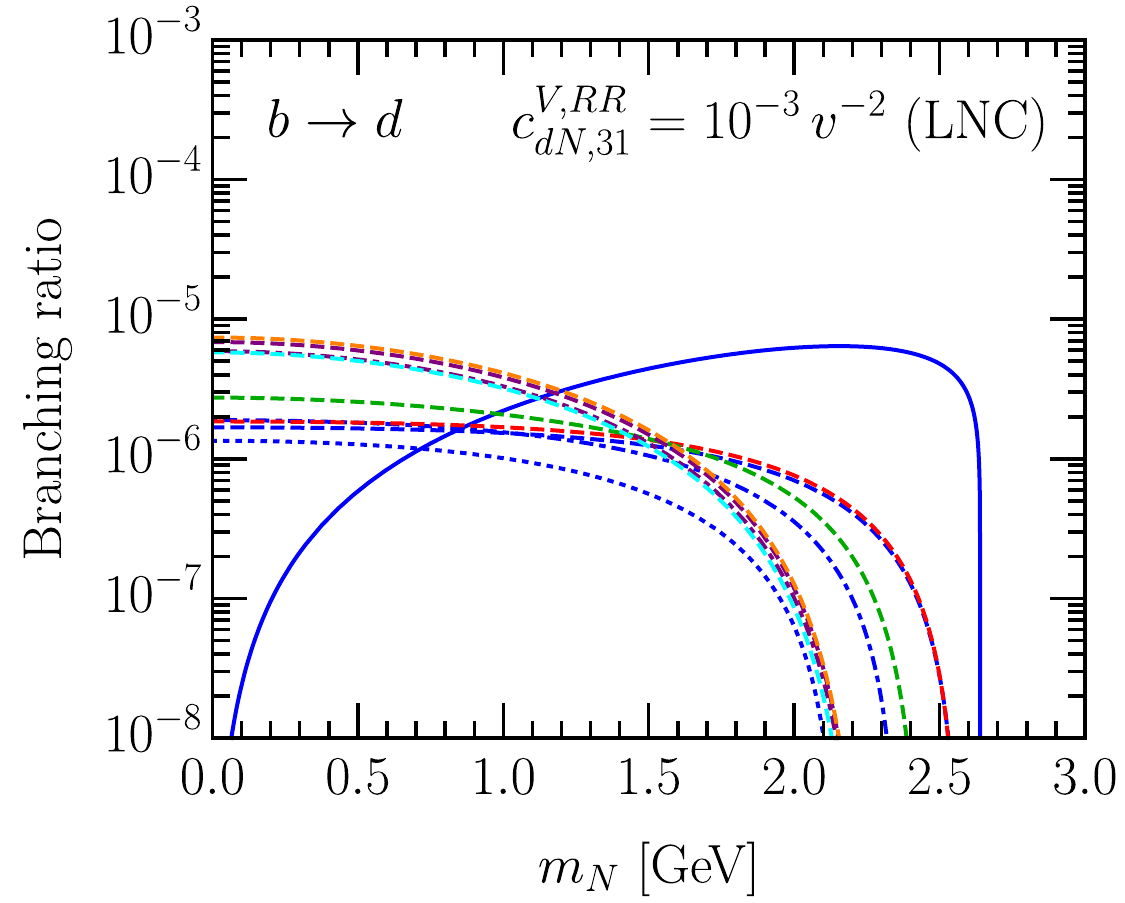} 
 \includegraphics[height=0.29\textheight]{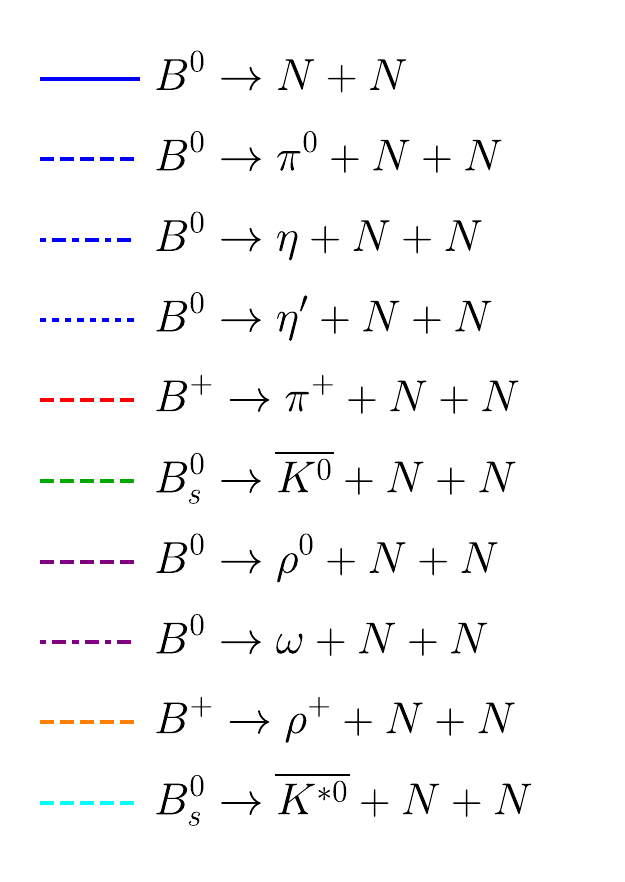}\\[0.15cm]
 \includegraphics[height=0.29\textheight]{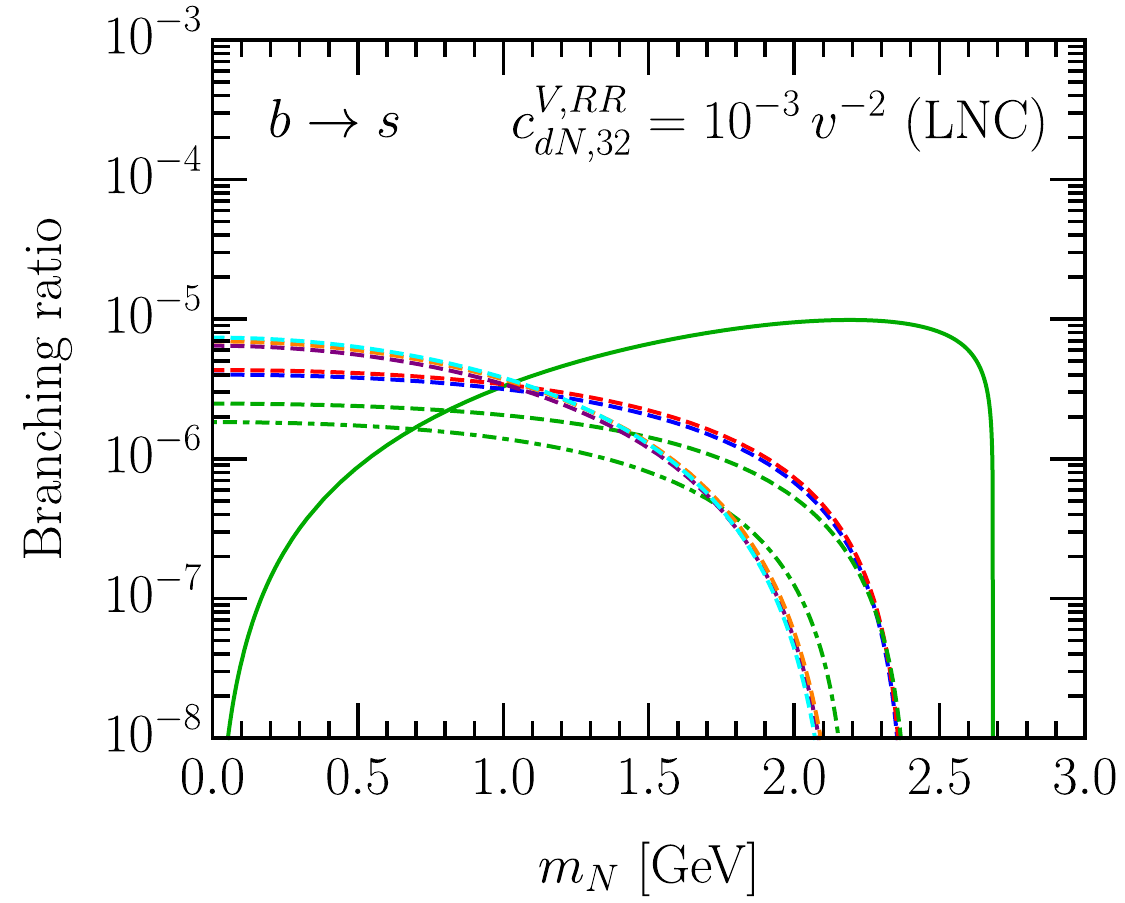}
 \includegraphics[height=0.29\textheight]{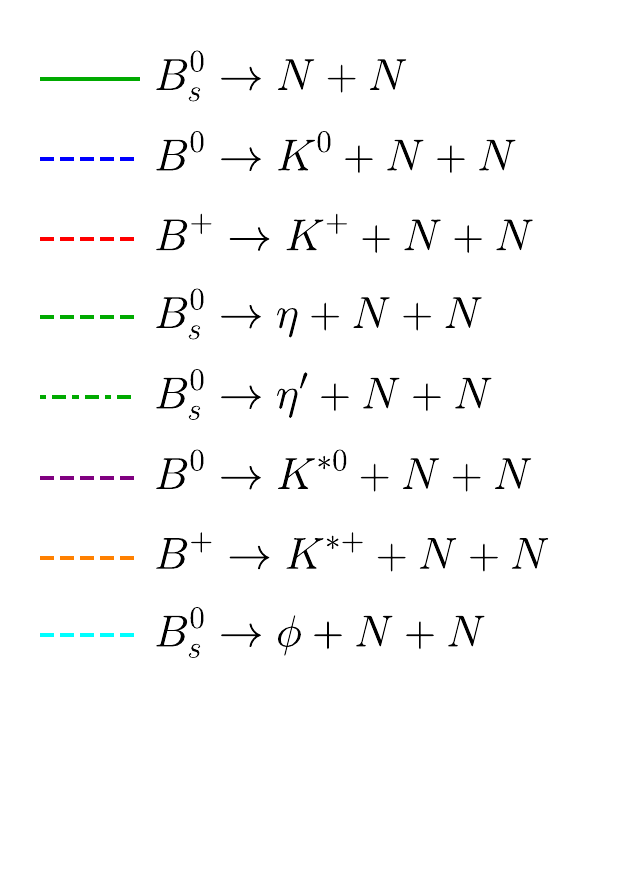}\\
 \caption{Branching ratios of $D$- and $B$-meson decays triggered 
by the LNC pair-$N_R$ operators $\mathcal{O}_{uN,12}^{V,RR}$ (top), 
$\mathcal{O}_{dN,31}^{V,RR}$ (middle) 
and $\mathcal{O}_{dN,32}^{V,RR}$ (bottom). 
The corresponding Wilson coefficient has been set to $10^{-3} v^{-2}$.}
 \label{fig:BrNNLNC}
\end{figure}

In figure~\ref{fig:BrNNLNC}, we show the branching ratios 
of $D$- and $B$-meson decays induced by the LNC pair-$N_R$ operators 
$\mathcal{O}_{uN,12}^{V,RR}$ (top), $\mathcal{O}_{dN,31}^{V,RR}$ (middle) 
and $\mathcal{O}_{dN,32}^{V,RR}$ (bottom), assuming $N$ is Majorana.
%
We turn on one operator at a time, setting the corresponding 
Wilson coefficient to $10^{-3} v^{-2}$. 
This choice allows us to avoid the bounds from (semi-)invisible meson decays 
summarized in table~\ref{tab:invisible}.
We see that the two-body decay dominates over the three-body ones 
for $m_N \gtrsim 325$~MeV in the case of $\mathcal{O}_{uN,12}^{V,RR}$, 
for $m_N \gtrsim 1.2$~GeV in the case of $\mathcal{O}_{dN,31}^{V,RR}$, 
and for $m_N \gtrsim 1.05$~GeV in the case of $\mathcal{O}_{dN,32}^{V,RR}$. 
For small $m_N$, the two-body decay is suppressed, 
since the corresponding decay width is proportional to $m_N^2$, see Eq.~\eqref{eq:GammaPNNMaj} in appendix~\ref{sec:decay_formulae}. 
Turning on the LR operators $\mathcal{O}_{uN,12}^{V,LR}$, 
$\mathcal{O}_{dN,31}^{V,LR}$, and $\mathcal{O}_{dN,32}^{V,LR}$ (one at a time) 
would lead to the same results as can be seen from 
Eqs.~\eqref{eq:GammaPNNMaj}, 
\eqref{eq:MPPpNNMaj} and \eqref{eq:MPVNNMaj}.
 
\begin{figure}[t]
 \centering
 \includegraphics[height=0.29\textheight]{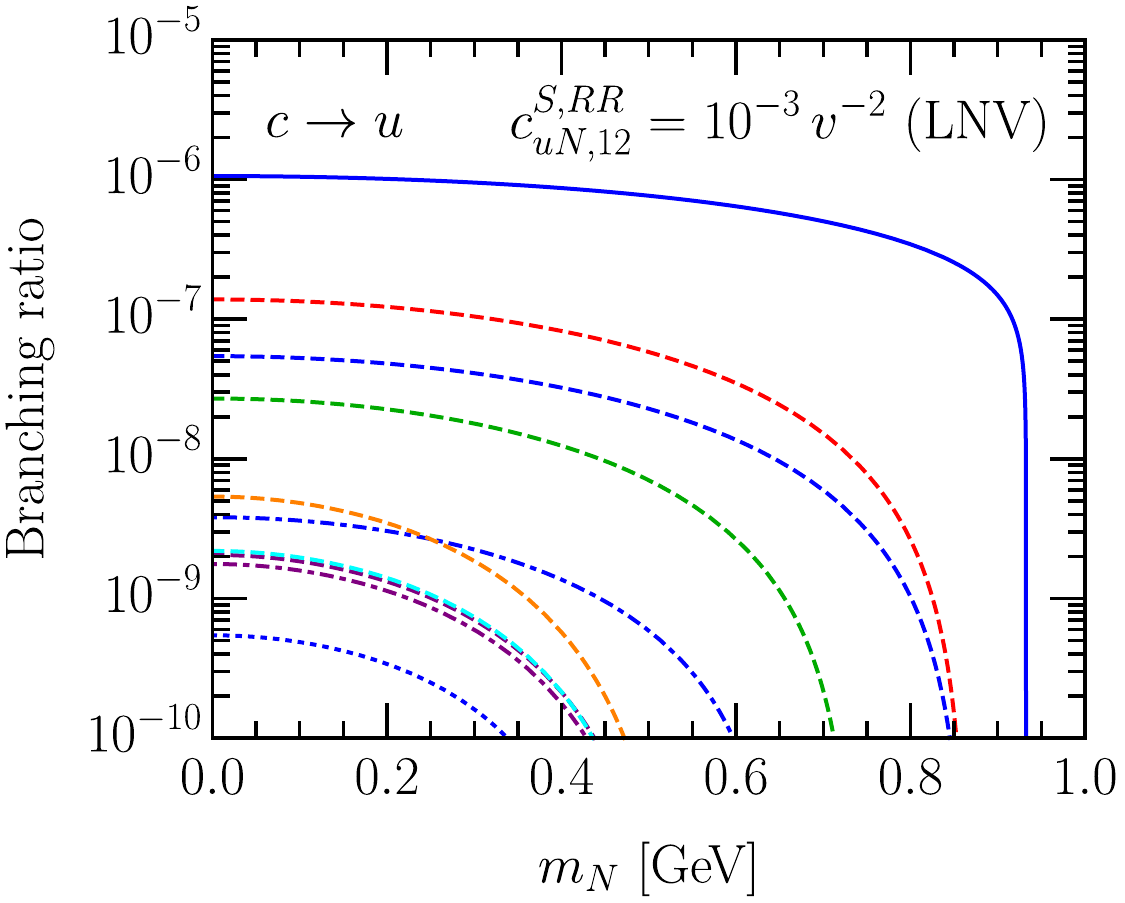}
 \includegraphics[height=0.29\textheight]{legend_D.pdf}\\[0.15cm]
 \includegraphics[height=0.29\textheight]{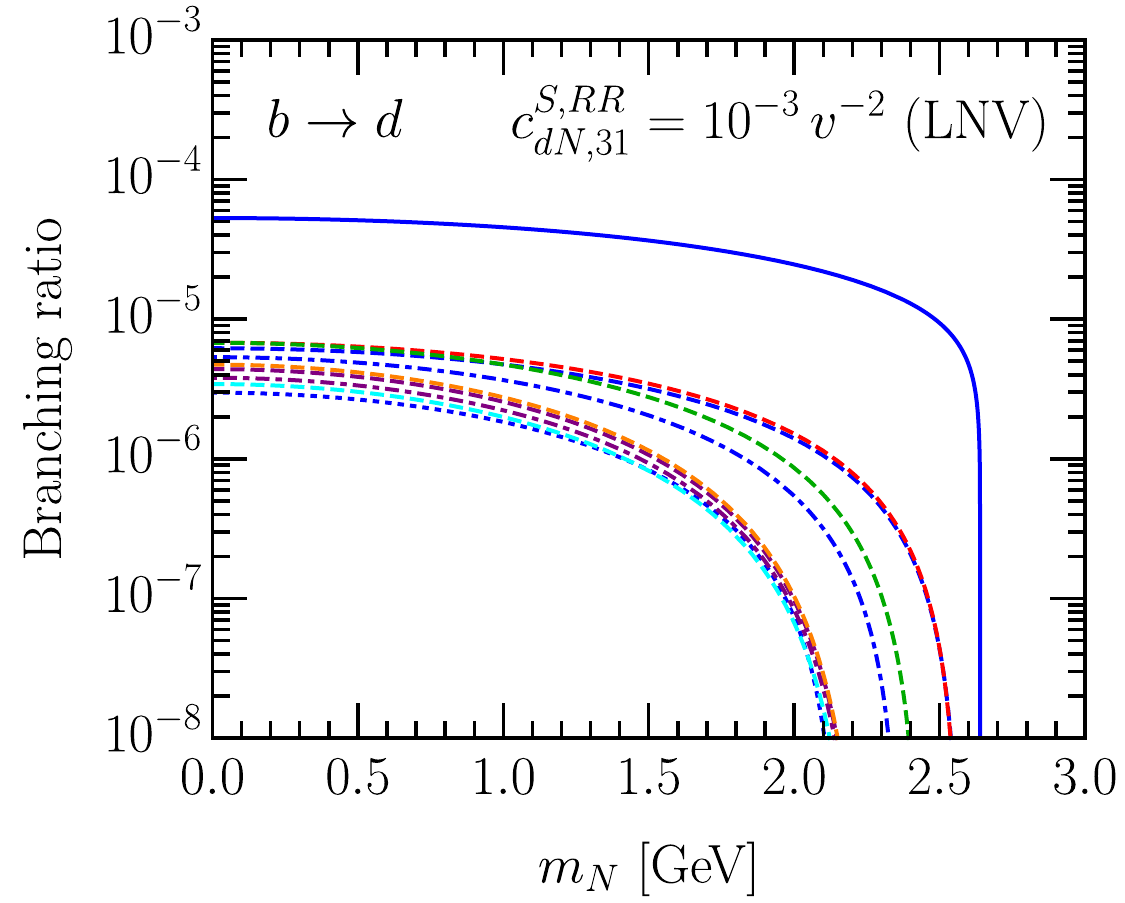} 
 \includegraphics[height=0.29\textheight]{legend_B.pdf}\\[0.15cm]
 \includegraphics[height=0.29\textheight]{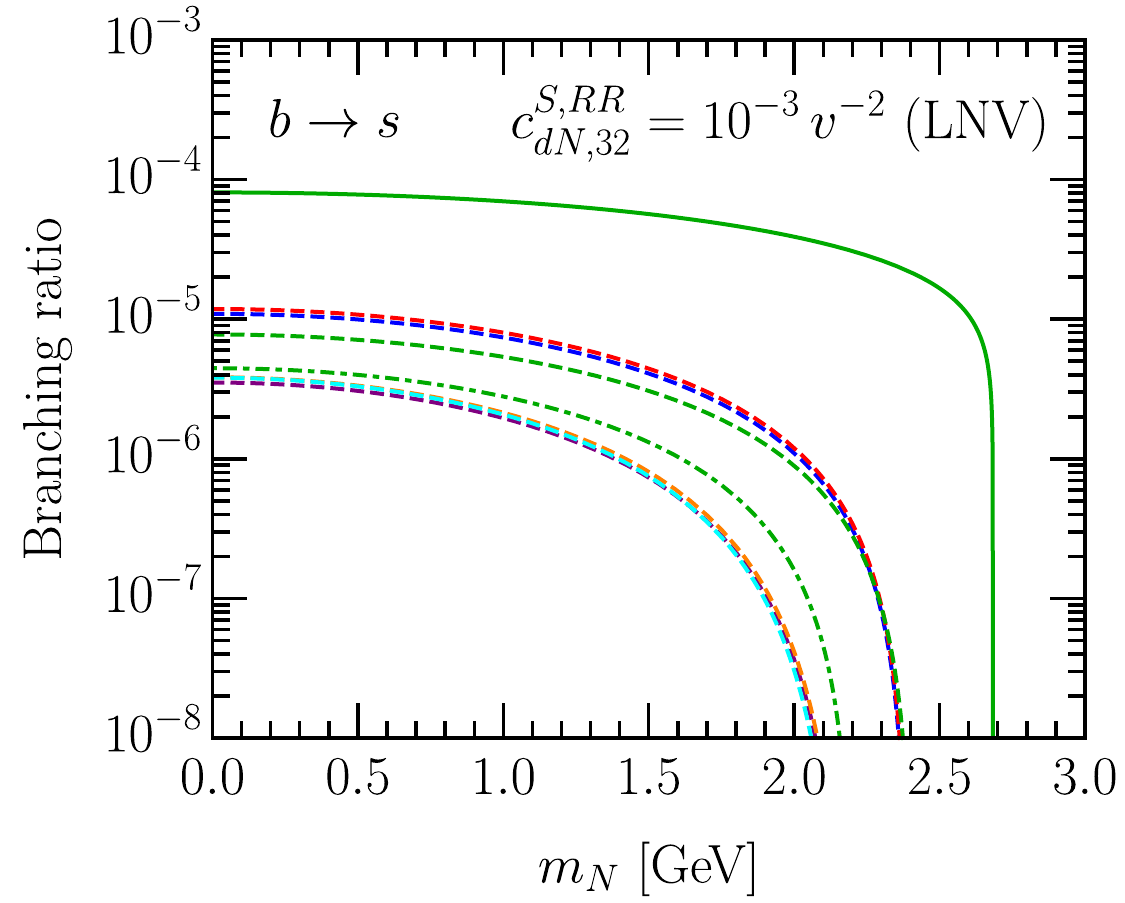}
 \includegraphics[height=0.29\textheight]{legend_Bs.pdf}\\
 \caption{Branching ratios of $D$- and $B$-meson decays triggered 
by the LNV pair-$N_R$ operators $\mathcal{O}_{uN,12}^{S,RR}$ (top), 
$\mathcal{O}_{dN,31}^{S,RR}$ (middle) 
and $\mathcal{O}_{dN,32}^{S,RR}$ (bottom). 
The corresponding Wilson coefficient has been set to $10^{-3} v^{-2}$.}
 \label{fig:BrNNLNV}
\end{figure}

In figure~\ref{fig:BrNNLNV}, we display the branching ratios 
of the same meson decays, but this time triggered 
by the LNV pair-$N_R$ operators 
$\mathcal{O}_{uN,12}^{S,RR}$ (top), 
$\mathcal{O}_{dN,31}^{S,RR}$ (middle), and $\mathcal{O}_{dN,32}^{S,RR}$ (bottom).
%
Again, we turn on one operator at a time and set the corresponding 
Wilson coefficient to $10^{-3} v^{-2}$.%
\footnote{In the case of non-hermitian operators, 
as \textit{e.g.} $\mathcal{O}_{qN}^{S,RR}$, $q=u$ or $d$, ``one operator at a time'' means 
$c_{qN,ij}^{S,RR} \neq 0$, whereas $c_{qN,ji}^{S,RR} = 0$.
We note also that $c_{dN,31}^{S,RR} = 10^{-3} v^{-2}$ violates slightly
the limit on $\mathrm{BR}(B^0 \to \text{inv.})$ from table~\ref{tab:invisible}. 
However, we opt for this value of the Wilson coefficient to be uniform with the other plots.} 
Contrary to the LNC operators, now the leptonic two-body decays 
dominate for all kinematically available values of $m_N$, 
since there is no helicity suppression from $m_N$, 
cf.~Eq.~\eqref{eq:GammaPNNMaj}. 
Switching on the LR operators 
$\mathcal{O}_{uN,12}^{S,LR}$, 
$\mathcal{O}_{dN,31}^{S,LR}$, 
and $\mathcal{O}_{dN,32}^{S,LR}$ (one at a time) 
would lead to the same results as can be inferred from 
Eqs.~\eqref{eq:GammaPNNMaj}, 
\eqref{eq:MPPpNNMaj}, and \eqref{eq:MPVNNMaj}. 
The tensor operators 
$\mathcal{O}_{dN}^{T,RR}$ and $\mathcal{O}_{uN}^{T,RR}$ 
vanish identically for one generation of $N_R$. 
For this reason, we do not consider them in what follows.

Concerning single-$N_R$ operators, we have also studied their contribution to two- and three-body $D$- and $B$- meson decays, showing some results in appendix~\ref{sec:BRnuN}. In particular, we present in figure~\ref{fig:BrBnuN} the branching ratios of $B$-meson decays induced by the LNC operators $\mathcal{O}_{d\nu N,31\alpha}^{S,RR}$ and
$\mathcal{O}_{d\nu N,31\alpha}^{T,RR}$ and the LNV operator
$\mathcal{O}_{d\nu N,31\alpha}^{V,RR}$. We observe that our results are analogous to the ones obtained in Ref.~\cite{DeVries:2020jbs}, cf.~figure~2 therein, for single-$N_R$ operators including a charged lepton. Therefore, we expect the constraints derived from the far detectors to be essentially the same for the neutral-current single-$N_R$ operators, and we hence do not include the latter in our numerical simulation.

Before closing the section, we mention that with only one $N_R$ generation, pair-$N_R$ operators cannot make the HNL decay by themselves, in contrast to single-$N_R$ operators. In appendix~\ref{sec:decaysingleN}, we have computed the partial decay widths of HNLs into a light meson and an active neutrino triggered by the neutral single-$N_R$ operators. Since we only simulate HNLs produced by pair-$N_R$ operators and only turn on one $N_R$LEFT operator at a time, we consider that the HNL decay is exclusively due to active-heavy neutrino mixing. The partial decay widths of HNL via mixing are computed according to the formulae given in Refs.~\cite{Atre:2009rg,Bondarenko:2018ptm,DeVries:2020jbs}.

\section{Experiments and details of simulation}
\label{sec:simulation}

In this work, we focus on a series of far-detector experiments proposed or even approved at the LHC.
These programs were mostly proposed in recent years and include  ANUBIS~\cite{Bauer:2019vqk}, AL3X~\cite{Gligorov:2018vkc}, CODEX-b~\cite{Gligorov:2017nwh}, FACET~\cite{Cerci:2021nlb}, FASER and FASER2~\cite{Feng:2017uoz,FASER:2018eoc}, MAPP1 and MAPP2~\cite{Pinfold:2019nqj,Pinfold:2019zwp}, and MATHUSLA~\cite{Chou:2016lxi,Curtin:2018mvb,MATHUSLA:2020uve}. 
Since they are supposed to be operated at different interaction points (IPs) and time scales of the LHC, the projected integrated luminosities also vary, ranging from as low as 30 fb$^{-1}$ for MAPP1, up to the full HL-LHC target, 3 ab$^{-1}$ for ANUBIS, FACET, FASER2, and MATHUSLA.
Moreover, typically these far detectors are located with a distance of $\mathcal{O}(1)-\mathcal{O}(100)$ m to their corresponding IP, allowing for sufficient amount of rock, lead, or other shielding material to effectively remove the potential SM background events stemming from the IP.
As a result, vanishing background is usually assumed for phenomenological studies on these experiments.

Here, we study HNLs pair-produced at the LHC with the center-of-mass energy 
$\sqrt{s}=14$ TeV from charm and bottom mesons' rare decays mediated by a $N_R$LEFT operator.
These HNLs then travel a certain distance before decaying via their mixing with the SM active neutrino.
If the boosted decay length of the HNLs falls into a suitable range, the HNLs have a high probability of decaying inside some of the far detectors.
Since the HNLs mostly decay to channels that are not fully invisible (three active neutrinos), the decay vertex can be reconstructed by these experiments.

While we analytically compute the production and decay rates of the HNLs, we rely on a Monte-Carlo simulation tool, Pythia8.3~\cite{Bierlich:2022pfr,Sjostrand:2014zea}, to estimate the decay probabilities of the HNLs inside each of the far detectors.
Depending on the flavor indices of the EFT operator we choose to turn on, Pythia8 can generate either the $pp\to c\bar{c}$ or $pp\to b\bar{b}$ process inclusively for $D$- or $B$-meson production at the LHC, respectively.
These mesons then decay to different channels including at least two HNLs, with relative decay branching ratios implemented according to the analytical computation detailed in section~\ref{sec:hnlproduction}. 
This allows for maximal usage of all the simulated data while respecting the relative importance of each decay channel.
Since Pythia8 provides the kinematics of each simulated HNL, we can compute its decay probability in each detector according to the geometry and position with respect to the IP.
We then take the average of the decay probabilities of all the simulated HNLs, to obtain the expected acceptance rate at each detector, $\langle P[N \text{ decay}] \rangle$.
This allows us to compute the projected number of signal events, $N_S$, with the following equation:
\begin{eqnarray}
N_S = \sum_{i} 2 \cdot  N_{M_i} \cdot \text{BR}(M_i \to N N + \text{anything}) \cdot \langle P[N \text{ decay}] \rangle \cdot \text{BR}(N \to \text{vis.}),
\end{eqnarray}
where the factor 2 accounts for the fact that each decay of the mesons into the HNLs via the $N_R$LEFT operator includes two HNLs, $N_{M_i}$ is the number of the mother mesons, and $\text{BR}(N \to \text{vis.})$ denotes the decay branching ratio of the HNLs into visible final states. 
For our study, we consider only the decay channel into three active neutrinos to be invisible.

In table~\ref{tab:mesonnumber}, we reproduce the inclusive production rates of all the relevant types of heavy mesons  from Ref.~\cite{DeVries:2020jbs}, which were obtained therein combining both LHCb measurements and extrapolation with the state-of-the-art simulation tools FONLL~\cite{Cacciari:1998it,Cacciari:2001td} and Pythia8.
Note that in table~\ref{tab:mesonnumber} we have in addition multiplied the production rates of these heavy mesons given in Ref.~\cite{DeVries:2020jbs} by a factor of 1.08 and 1.06 for the $B$- and $D$-mesons, respectively, in order to take into account the fact that we consider $\sqrt{s}=14$ TeV instead of 13 TeV~\cite{Hirsch:2020klk}.
\begin{table}[t]
\centering 
 \begin{tabular}{|c|c|c|c|c|c|}
 \hline
	$D^0$					&$D^\pm$                        & $D^\pm_s$
	&$B^0$					&$B^\pm$    	               & $B_s^0$ 		\\
	$4.12\times 10^{16}$	& $2.16\times 10^{16}$    & $7.02\times 10^{15}$   &$1.58 \times10^{15}$		& $1.58\times 10^{15}$     & $2.73\times 10^{14}$   \\
 \hline
 \end{tabular}
 \caption{Inclusive production numbers of $D$- and $B$-mesons for an integrated luminosity of 3 ab$^{-1}$, extracted from Ref.~\cite{DeVries:2020jbs}, but multiplied by a factor of 1.08 (1.06) for the $B$-mesons ($D$-mesons) obtained with Pythia8 to account for the increase in production cross section of the heavy mesons from the center-of-mass energy 13~TeV to 14~TeV~\cite{Hirsch:2020klk}.}
 \label{tab:mesonnumber}
\end{table}
%

In principle, the computation of the decay probability of each HNL can be summarized with a simple formula, if the flight path of the HNL traverses the detector:
\begin{eqnarray}\label{eq:decayprob}
P[N \text{ decay}] = \text{e}^{-L_1/\beta\gamma c\tau} -  \text{e}^{-L_2/\beta\gamma c\tau},
\end{eqnarray}
where $\beta$, $\gamma$, and $\tau$ are the speed, Lorentz boost factor, and the proper lifetime of the HNL, $c$ is the speed of light, and $L_1$ and $L_2$ label respectively the distance from the IP (where we assume the heavy mesons decay into the HNLs instantly) to the incoming and outgoing sides of the detector.
This formula holds perfectly for a far detector with a spherical-shell shape facing towards the IP. 
However, since the far-detector experiments mostly have a more complicated geometrical shape and relative orientation with respect to the IP, more sophisticated formulae are required for more accurate estimate of the decay probability.
In this paper, we strictly follow Ref.~\cite{DeVries:2020jbs} for computing the decay probabilities in each far detector with the corresponding geometrical configuration.
As exception, FACET~\cite{Cerci:2021nlb} was not included in Ref.~\cite{DeVries:2020jbs}, so here, we give a brief description 
of this detector.
Its concept is essentially rather similar to that of FASER(2) and AL3X.
It is a proposed cylindrical detector in the forward direction of the CMS experiment, with a distance of 101 m between the CMS IP and its near end.
Further, it has a radius of 0.5 m and length 18 m.
However, it does not cover the pseudorapidity up to infinity as FASER(2) does; instead, it covers the polar angle between 1 and 4 mrad, or equivalently, the pseudorapidity between 6.2 and 7.6.
We require that the HNLs travel towards the far end of the FACET decay chamber (but not to the sides).

Before closing, we should also mention that since Pythia8 is not well validated for  
heavy meson production in the very forward direction, we use FONLL to correct the differential production cross section of mesons  
for various ranges of small transverse momentum and large pseudorapidity, for three experiments: FACET, FASER, and FASER2. 

\section{Results of numerical analysis}
\label{sec:results}

\begin{figure}[t]
 \centering
 \includegraphics[width=\textwidth]{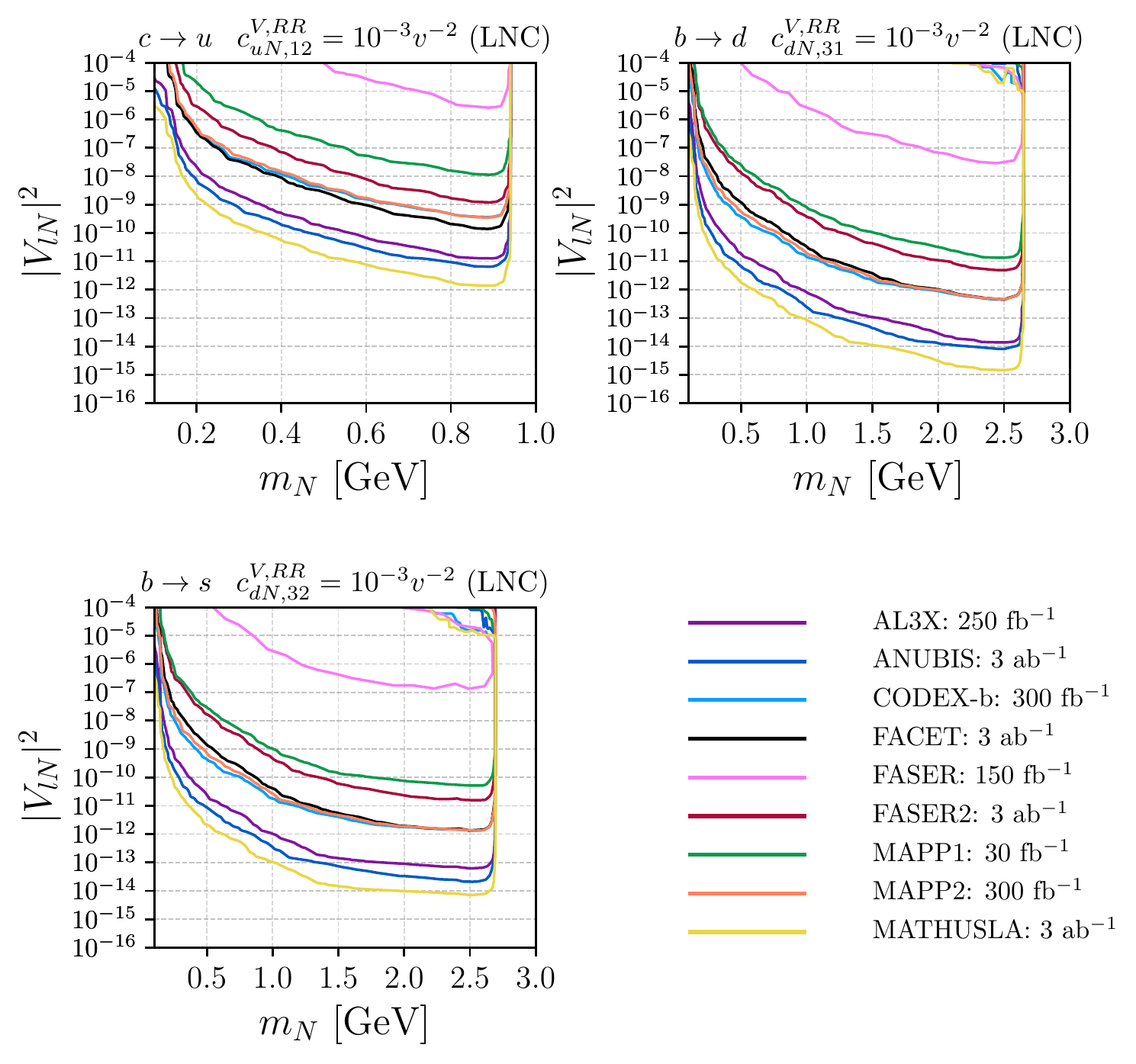}
 \caption{Exclusion limits in the plane $|V_{lN}|^2$ vs.~$m_N$, for the three LNC pair-$N_R$ operators we consider, with the corresponding Wilson coefficients fixed to $10^{-3} v^{-2}$.}
 \label{fig:vv_vs_mass_LNC}
\end{figure}

\begin{figure}[t]
 \centering
 \includegraphics[width=\textwidth]{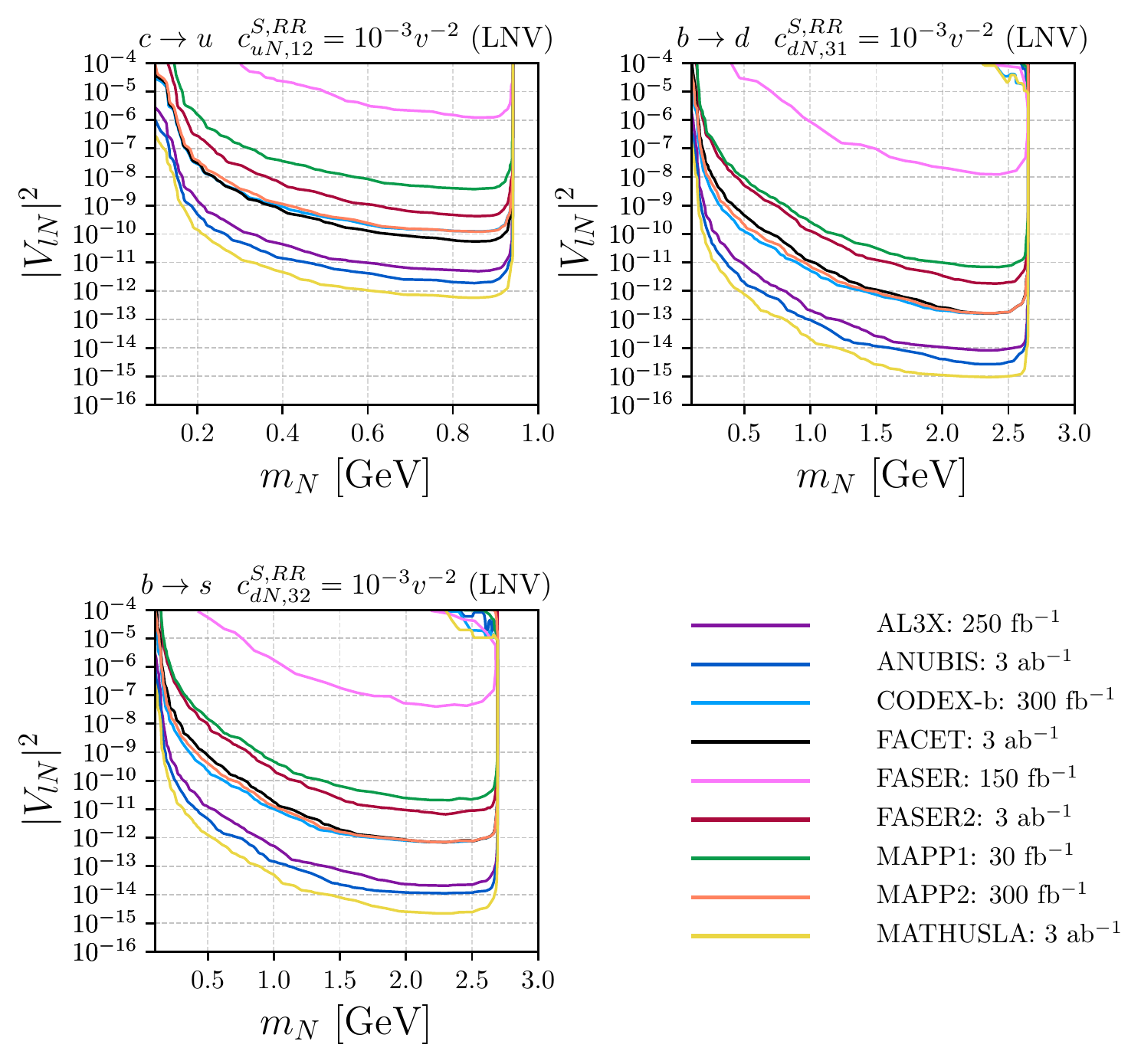}
 \caption{The same plots as Fig.~\ref{fig:vv_vs_mass_LNC}, but for the three LNV pair-$N_R$ operators.}
 \label{fig:vv_vs_mass_LNV}
\end{figure}

We present and discuss the numerical results in this section.
We restrict ourselves to Majorana HNLs, and study 
the LNC and LNV pair-$N_R$ operators in the $N_R$LEFT 
triggering the meson decays 
described in section~\ref{sec:hnlproduction}. More specifically, 
the set of the LNC operators we consider is 
$\mathcal{O}_{uN,12}^{V,RR}$, $\mathcal{O}_{dN,31}^{V,RR}$, and 
$\mathcal{O}_{dN,32}^{V,RR}$, 
and the set of the LNV operators is 
$\mathcal{O}_{uN,12}^{S,RR}$, $\mathcal{O}_{dN,31}^{S,RR}$, and 
$\mathcal{O}_{dN,32}^{S,RR}$.%
\footnote{We recall that the corresponding LR operators, 
if switched on one at a time, would lead to the same results.}
We first show in figures~\ref{fig:vv_vs_mass_LNC} and \ref{fig:vv_vs_mass_LNV} sensitivity reaches of the various far detectors we consider, in the plane $|V_{lN}|^2$ vs.~$m_N$, for the LNC and LNV operators, respectively.
These isocurves are for three signal events, corresponding to 95\% confidence level (C.L.) limits, given zero background events.
We switch on one operator only each time and fix the corresponding Wilson coefficient to $10^{-3}v^{-2}$, where $v=246$ GeV is the SM Higgs VEV.
This is approximately 700 times smaller than the Fermi constant and, for the LNC operators, corresponds to a NP scale $\Lambda_\mathrm{NP}$ of 
7.78~TeV for $c\sim 1/\Lambda_\mathrm{NP}^2$.
For this value of the Wilson coefficient, the limits coming 
from the $D^0 $ and $B^0$ invisible decays, 
as well as from $P \to P'/V \nu \overline{\nu}$, 
are avoided (see section~\ref{sec:hnlproduction}).
In our numerical simulation, we consider the HNLs only dominantly mixing with either the electron neutrino or the muon neutrino: $l=e,\mu$. 
We do not consider HNLs decaying to $\tau$'s,
because we expect the sensitivities to be much reduced, compared to $e,\mu$.
%
%
%

In figure~\ref{fig:vv_vs_mass_LNC}, we observe that in the 
top-right and bottom-left plots, there is stronger reach in $m_N$ than in the top-left plot.
This is due to the larger masses of $B$-mesons than those of $D$-mesons and hence higher kinematic thresholds.
Moreover, the $B$-meson scenarios show better exclusion limits in $|V_{lN}|^2$ as well, compared to the $D$-meson one, despite the larger production rates of $D$-mesons than those of $B$-mesons by about one order of magnitude, cf.~table~\ref{tab:mesonnumber}.
The main reason is that the decay branching ratios of $B$-mesons into the HNLs are in general larger than those of $D$-mesons by about 
two orders of magnitude, cf.~figure~\ref{fig:BrNNLNC}, assuming equal values of the corresponding Wilson coefficients.

Among these far-detector experiments, we find the best limits come from MATHUSLA, 
which can probe the values of $|V_{lN}|^2$ down to $10^{-12}$ ($10^{-15}$) 
in the charm (bottom) scenario(s), 
followed by ANUBIS and AL3X, while FASER shows the weakest sensitivities, 
mainly because of its small volume.
We note that the approved experiment MAPP1 with only 30~fb$^{-1}$ of integrated luminosity can already compete with the future experiment FASER2 with the full HL-LHC data, especially in the bottom scenarios. 
The reason is its larger size and smaller distance to the corresponding IP.
Also, CODEX-b, MAPP2, and FACET can exclude almost the same parts of the parameter space.

The plots in figure~\ref{fig:vv_vs_mass_LNV} are very similar to those in figure~\ref{fig:vv_vs_mass_LNC}, with the main difference being that the exclusion limits in $|V_{lN}|^2$ are slightly stronger because the decay branching ratios of the mesons into a pair of HNLs are larger in the LNV case than in the LNC case, given the same value of the Wilson coefficients, see figures~\ref{fig:BrNNLNC} and \ref{fig:BrNNLNV}.

The best present limits for HNL masses below 5 GeV that can be found in the literature come from different experiments such as NA62~\cite{NA62:2020mcv}, T2K~\cite{T2K:2019jwa}, CHARM~\cite{Bergsma:1985is}, PS191~\cite{Bernardi:1987ek}, JINR~\cite{Baranov:1992vq}, Belle~\cite{Liventsev:2013zz}, and DELPHI~\cite{Abreu:1996pa}. These limits are around $|V_{lN}|^2 \sim 10^{-9}$--$10^{-5}$, depending on the HNL mass, and correspond to a minimal HNL scenario in which both the production and the decay of the HNL are determined by their mixing with the active neutrinos. In our theoretical setup, however, the production of the HNLs receives additional contributions from effective operators affecting the HNL sensitivity of these experiments. A detailed reinterpretation of the current HNL limits coming from different experiments in the $N_R$LEFT depends on specific details of each particular experiment, and it is beyond the scope of this current work.
For this reason, we do not show these bounds in figures~\ref{fig:vv_vs_mass_LNC} and \ref{fig:vv_vs_mass_LNV}.

\begin{figure}[t]
	\centering
	\includegraphics[width=1.0\textwidth]{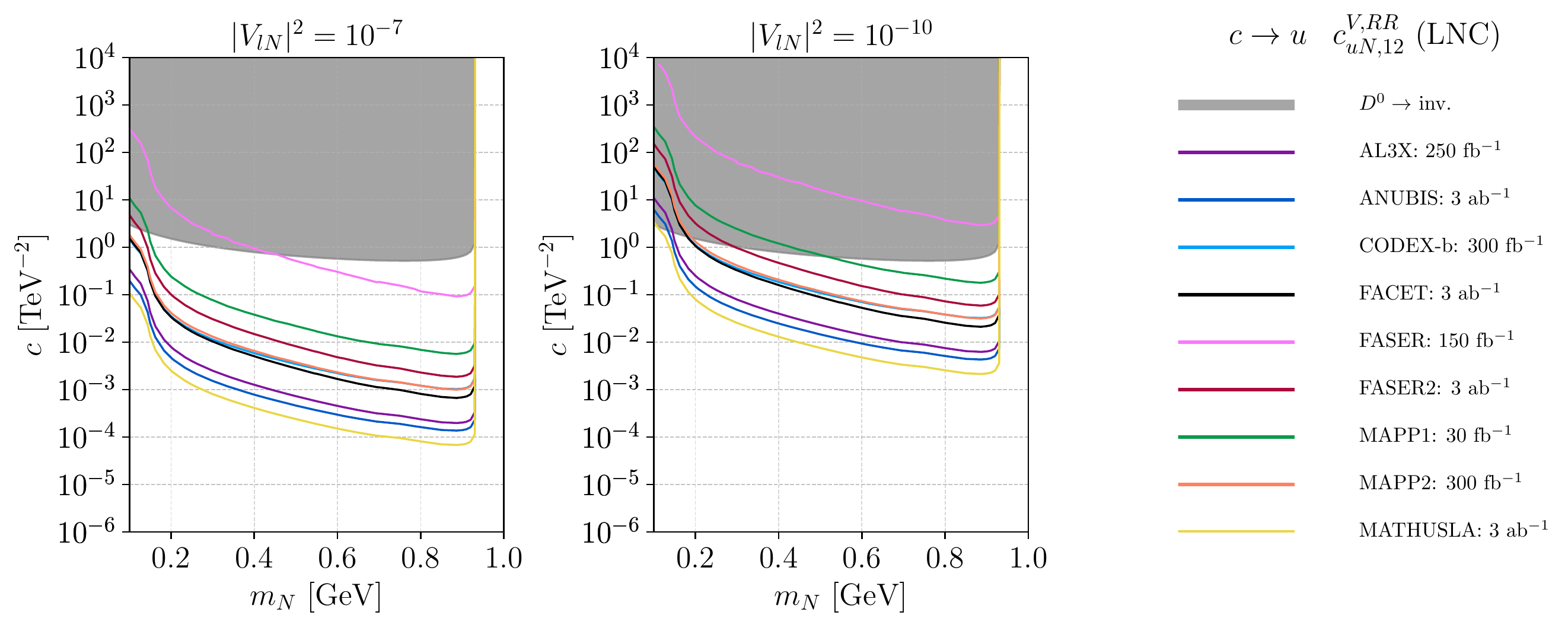}
	\includegraphics[width=1.0\textwidth]{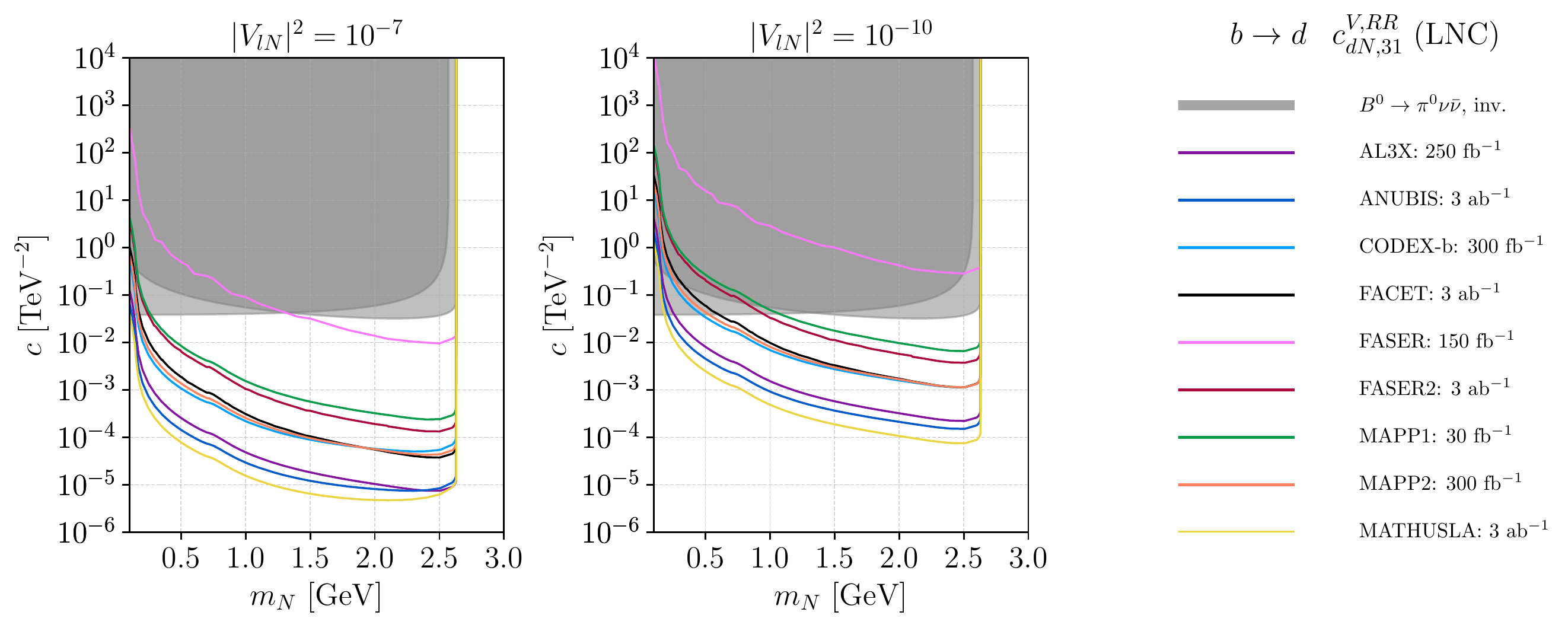}\\
	\includegraphics[width=1.0\textwidth]{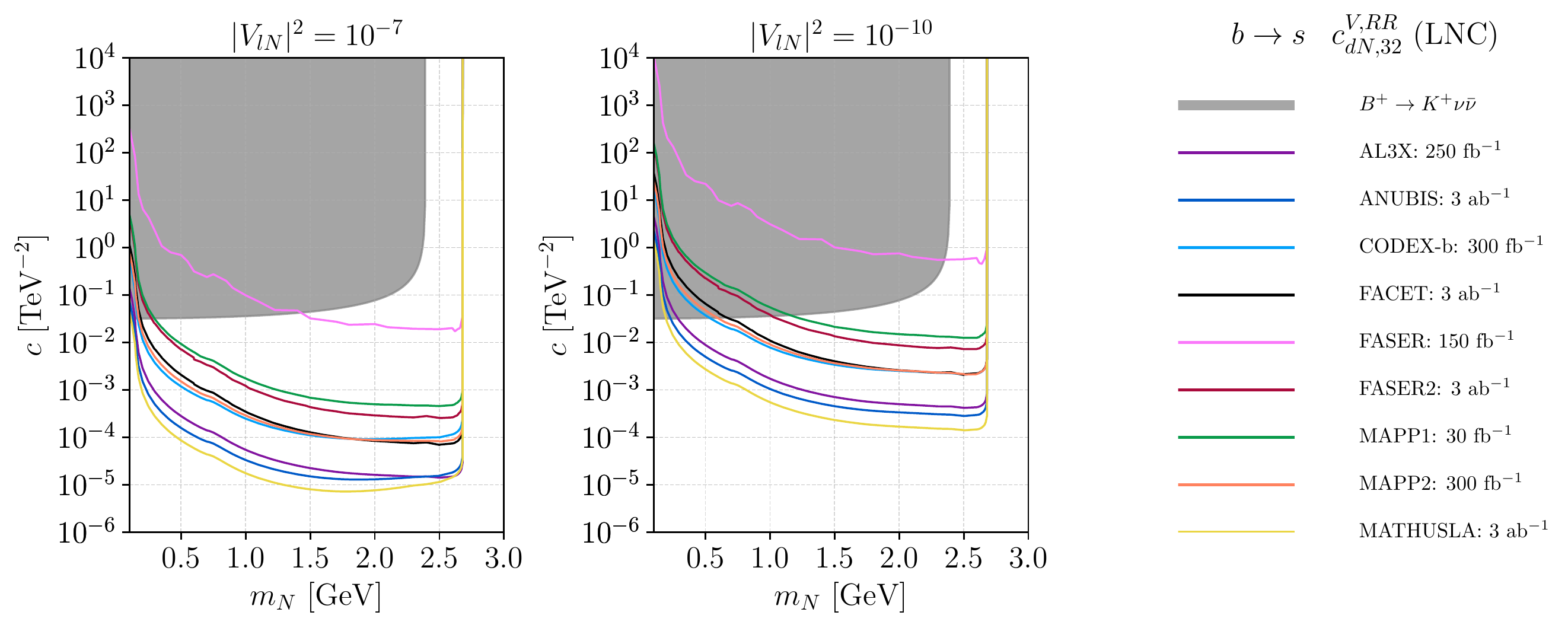}
	\caption{Exclusion limits in the plane $c$ vs.~$m_N$, for the three LNC pair-$N_R$ operators we consider, with $|V_{lN}|^2$ fixed at $10^{-7}$ or $10^{-10}$.
	}
	\label{fig:Lambda_vs_mass_LNC}
\end{figure}
%
%
\begin{figure}[t]
\centering
	\includegraphics[width=1.0\textwidth]{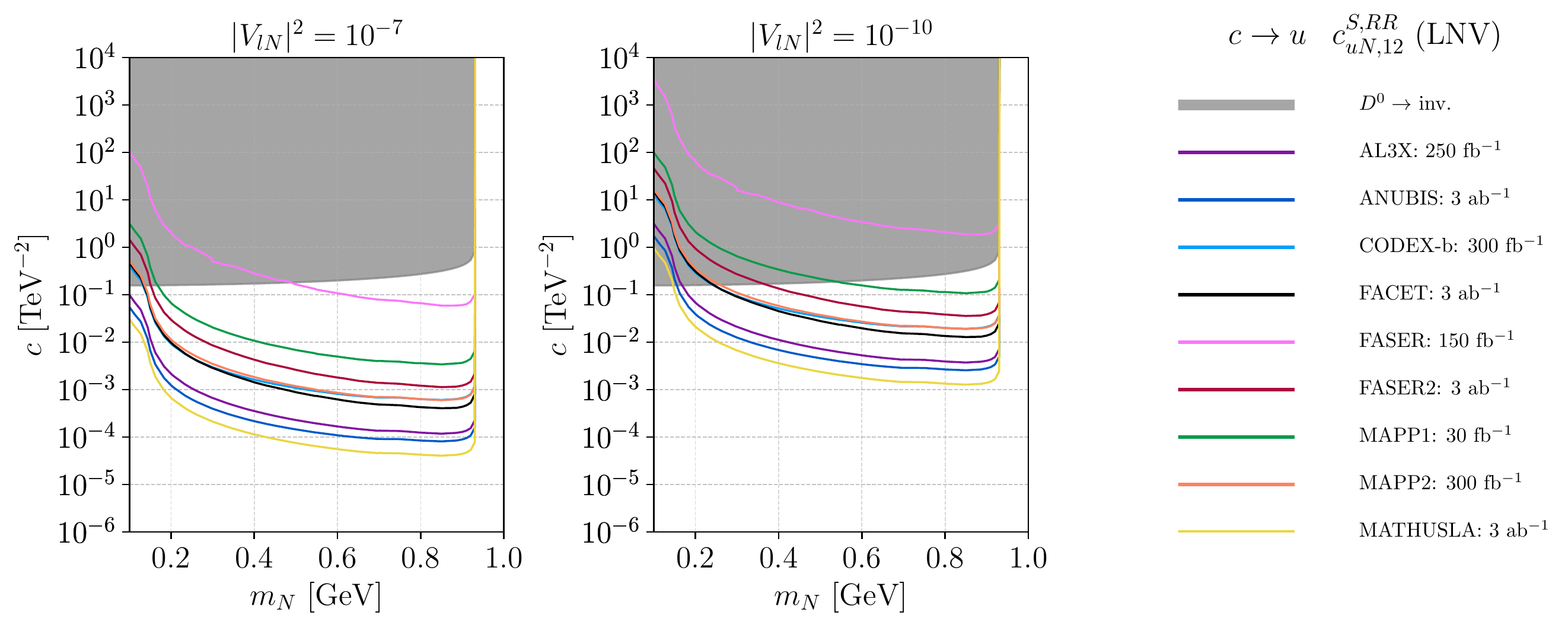}
	\includegraphics[width=1.0\textwidth]{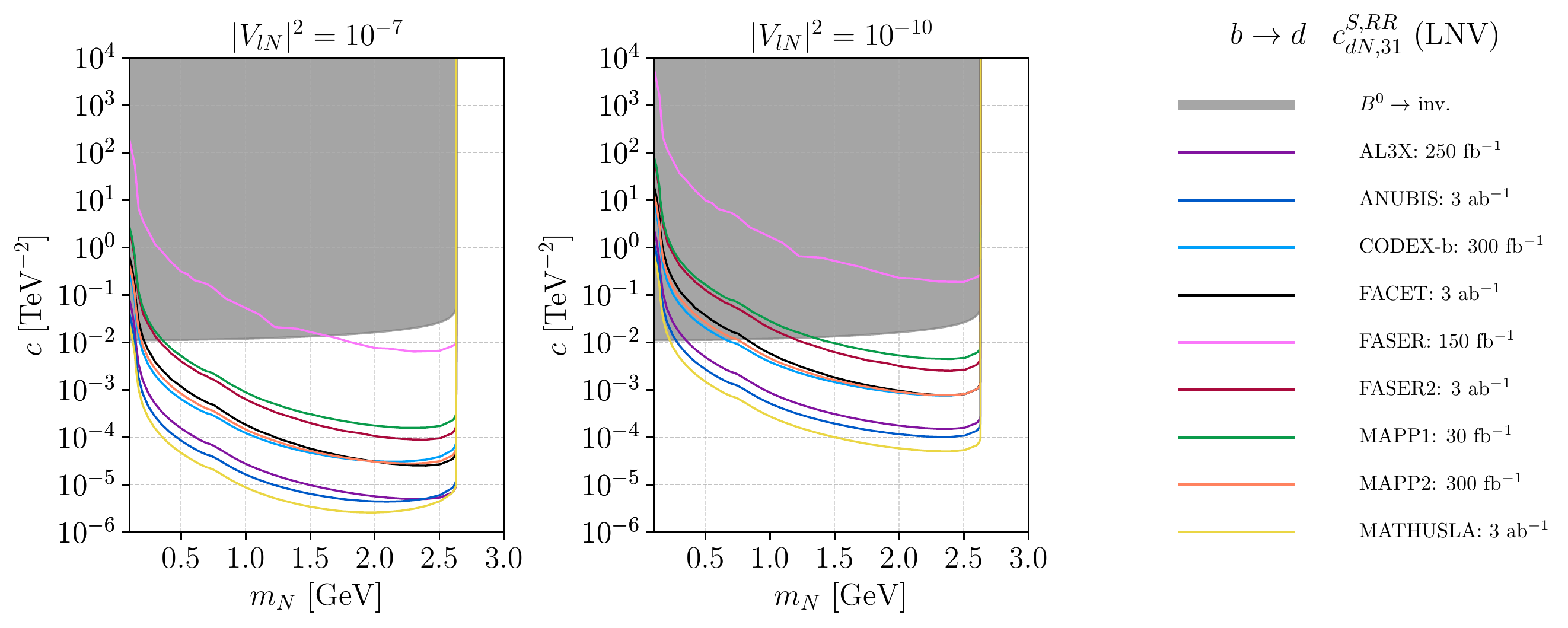}\\
	\includegraphics[width=1.0\textwidth]{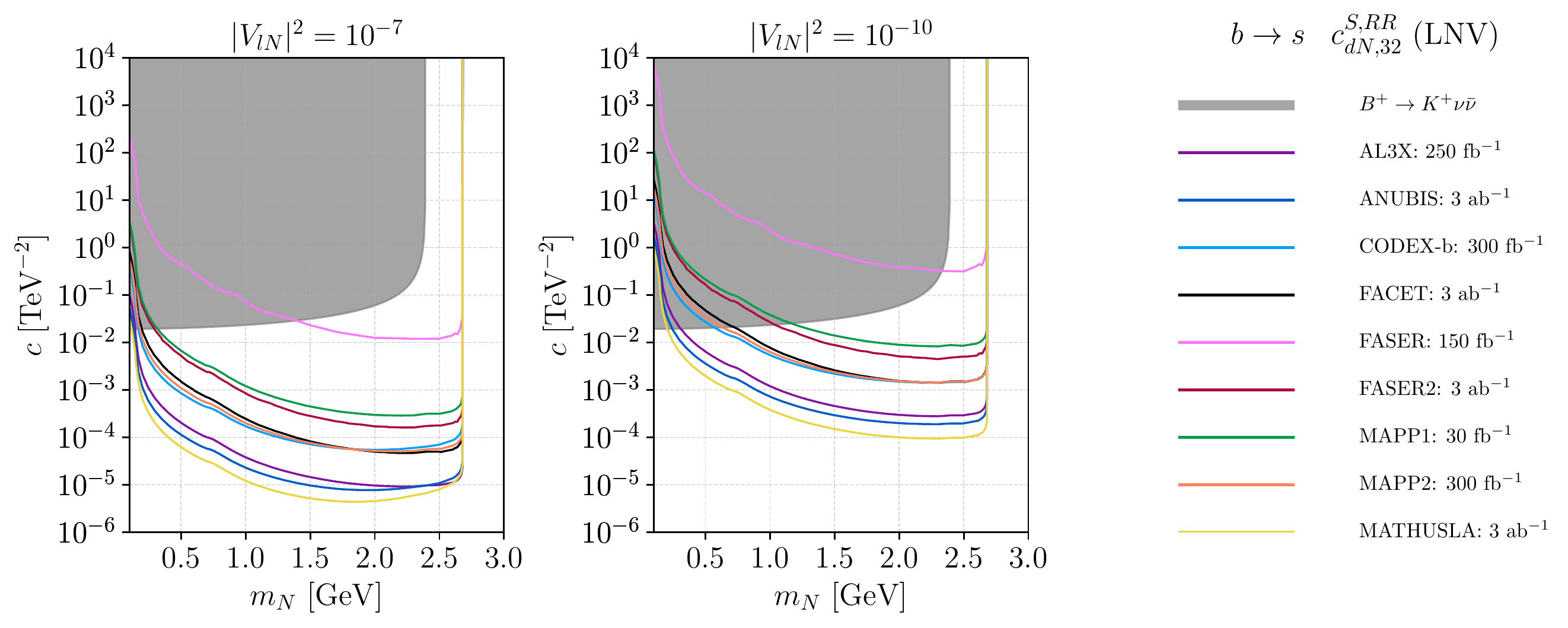}
\caption{The same plots as Fig.~\ref{fig:Lambda_vs_mass_LNC}, but for the three LNV pair-$N_R$ operators.}
\label{fig:Lambda_vs_mass_LNV}
\end{figure}
%
We then present in figures~\ref{fig:Lambda_vs_mass_LNC} and \ref{fig:Lambda_vs_mass_LNV} 
the sensitivity curves in the plane $c$ vs.~$m_N$, for two fixed values 
of active-heavy mixing, $|V_{lN}|^2 = 10^{-7}$ and $|V_{lN}|^2 = 10^{-10}$.
In order to obtain these plots, we select the chosen benchmark mixing parameters and find the value of the Wilson coefficient in each scenario 
such that the predicted number of signal events is equal to three.
We find the same hierarchy in sensitivities of these experiments 
as in figures~\ref{fig:vv_vs_mass_LNC} and \ref{fig:vv_vs_mass_LNV}, but the final reaches in $c$ differ between the two chosen $|V_{lN}|^2$ values.
This is because for any fixed HNL mass, varying the mixing parameter squared would change the proper lifetime of the HNL and hence the detectors' acceptance to these HNLs.
$|V_{lN}|^2=10^{-7}$ corresponds to an acceptance rate roughly three orders of magnitude better than that given by $|V_{lN}|^2=10^{-10}$, in the large decay length limit ($\beta\gamma c\tau \gg L_1, L_2$, as in Eq.~\eqref{eq:decayprob}).
Therefore, the lower sensitivity reach in $c$ for $|V_{lN}|^2=10^{-10}$ is roughly weaker than that for $|V_{lN}|^2=10^{-7}$, by a factor of $\sqrt{1000}\approx 32$, since the HNL production rates are proportional to $c^2$.
In the charm scenario, for the larger mixing parameter case, we find the best experiments (MATHUSLA, ANUBIS, and AL3X) can reach $c$ as low as about $10^{-4}$ TeV$^{-2}$, while the other weaker experiments can still probe $c$ down to the order of $10^{-3}$ TeV$^{-2}$ except FASER, or even better by about one order of magnitude in the two bottom scenarios.
In the LNC case, $c\sim 10^{-4}~(10^{-5})$ TeV$^{-2}$ corresponds roughly to a new physics scale $\Lambda_\mathrm{NP}$ of 100 (316) TeV, if we take $c\sim 1/\Lambda_\mathrm{NP}^2$ and ignore the relatively minor effects of QCD running.

Again as in figures~\ref{fig:vv_vs_mass_LNC} and \ref{fig:vv_vs_mass_LNV}, we find the LNV results in figure~\ref{fig:Lambda_vs_mass_LNV} are slightly stronger than the LNC ones in figure~\ref{fig:Lambda_vs_mass_LNC}, for the same reason as stated above.
Here, the conversion from Wilson coefficient to $\Lambda_\mathrm{NP}$ is more subtle than the LNC operators' case, because the $d=6$ LNV operators at the scale below 
$v=246$~GeV are matched to the $d=7$ operators 
that include a Higgs boson in the $N_R$SMEFT, cf.~Eq.~\eqref{eq:matching_LNV_pairN}.
Therefore, we estimate that for $c\sim 10^{-4}~(10^{-5})$ TeV$^{-2}$ in the low energy scale, the corresponding $\Lambda_\mathrm{NP}$ is about 10~(21)~TeV, with $c\sim v/(2\sqrt{2}\Lambda_\mathrm{NP}^3)$, if we again ignore the QCD running effects. This is weaker than the LNC case by about one order of magnitude.

The gray shaded regions in figures~\ref{fig:Lambda_vs_mass_LNC} 
and \ref{fig:Lambda_vs_mass_LNV} correspond to the limits coming from invisible 
and semi-invisible $D$- and $B$-meson decays collected in table~\ref{tab:invisible}. 
More specifically, in the $c \to u$ scenario, it is $D^0 \to \text{inv.}$~which provides the only constraint. 
In the $b \to d$ scenario and for the LNC operator (middle-row plots in figure~\ref{fig:Lambda_vs_mass_LNC}), there is an interplay between 
$B^0 \to \pi^0 \nu \overline{\nu}$, which dominates for $m_N \lesssim 1.3$~GeV, 
and $B^0 \to \text{inv.}$, which takes over for larger HNL masses.
For the LNV operator (figure~\ref{fig:Lambda_vs_mass_LNV}), 
$B^0 \to \text{inv.}$~provides the most stringent constraint 
for any kinematically allowed $m_N$. 
Finally, in the $b \to s$ scenario, the leading bound originates from the limit on 
the branching ratio of $B^+ \to K^+ \nu \overline{\nu}$. 
All these bounds have been derived by equating 
$\mathrm{BR}(P \to NN)$ ($\mathrm{BR}(P \to P' N N)$)
obtained using Eq.~\eqref{eq:GammaPNNMaj} (Eq.~\eqref{eq:MPPpNNMaj})
to the corresponding upper limit on 
$\mathrm{BR}(P \to \text{inv.})$ ($\mathrm{BR}(P \to P' \nu \overline{\nu})$) 
from table~\ref{tab:invisible}.


\section{Summary and conclusions}
\label{sec:summary}

In this work, we have focused on dimension-6 operators involving a pair of heavy neutral leptons (HNLs), in the framework of the low-energy effective field theory of the Standard Model extended with HNLs ($N_R$LEFT).
We have considered both lepton-number-conserving (LNC) and lepton-number-violating (LNV) operators.
Such operators with certain quark-flavor combinations can lead to charm and bottom meson decays into a pair of HNLs, in either two-body or three-body decays.
We have considered three combinations of the quark flavor indices, leading to the 
$c\to u$, $b\to d$, and $b\to s$ transitions, and computed the corresponding production rates of the HNL pair from various $D$- and $B$-meson decays for 
(i)~the LNC operators 
$\mathcal{O}_{uN,12}^{V,RR}$, $\mathcal{O}_{dN,31}^{V,RR}$, and 
$\mathcal{O}_{dN,32}^{V,RR}$, and 
(ii)~the LNV operators
$\mathcal{O}_{uN,12}^{S,RR}$, $\mathcal{O}_{dN,31}^{S,RR}$, and 
$\mathcal{O}_{dN,32}^{S,RR}$.
The corresponding decay branching ratios for both LNC and LNV cases 
are shown in figures~\ref{fig:BrNNLNC} and \ref{fig:BrNNLNV}, respectively.
Switching on the corresponding LR operators (one at a time) would lead to the same results.
It is interesting to observe that for the LNC case, the mesons' three-body and two-body decays are dominant, respectively, for relatively lower and higher masses in the kinematically allowed range of the HNL mass, for all three quark-flavor scenarios.
On the other hand, for the LNV case, the two-body decays are the most important channel for the whole kinematically relevant mass range.

Since these operators entail a pair of HNLs and 
in the absence of other effective interactions, HNLs can decay only via an additional HNL mixing with the active neutrinos in the weak current.%
\footnote{Actually, an HNL can also decay 
through a pair-$N_R$ operator in conjunction with mixing, but such decays are 
suppressed compared to the decays through mixing alone.}
Therefore, we study HNL production from the EFT operators, and 
HNL decays through active-heavy neutrino mixing.
Since we focus on GeV-scale HNLs, small neutrino mixing parameters naturally lead these HNLs to become long-lived.
Here, we choose therefore to consider a series of far-detector proposals at the LHC.
These include AL3X, ANUBIS, CODEX-b, FACET, FASER and FASER2, MAPP1 and MAPP2, and MATHUSLA.
These experiments have been proposed (some even approved and operated) mainly for searching for long-lived particles (LLPs) and would be constructed with a distance of about $5$--$500$ meters away from different interaction points (IPs).
With shielding materials on site between a far detector and its corresponding IP, such experiments are usually assumed to be background-free for LLP searches.

We then perform Monte-Carlo simulation with the tool Pythia8, to estimate the acceptance rates of these far detectors for HNLs pair-produced from $D$- and $B$-meson decays.
For small mixing and mass, these HNLs are long-lived and decay outside the LHC main detectors.
With the total production rates of the mesons at the LHC in knowledge, we can estimate the number of HNL decays in the far detectors, using decay width formulae for 
(i)~the HNL production from meson decays in the $N_R$LEFT, and 
(ii) the HNL decays via their mixing with the active neutrinos.
For simplicity, we consider the HNLs to be mixed with only either the electron neutrino or the muon neutrino.
Further, for the signal, we consider only the HNL decays into three active neutrinos to be invisible; all the other decay channels are included as signature.

We show our numerical results in two types of planes: $|V_{lN}|^2$ vs.~$m_N$ 
in figures~\ref{fig:vv_vs_mass_LNC} and \ref{fig:vv_vs_mass_LNV},
and $c$ vs.~$m_N$ in figures~\ref{fig:Lambda_vs_mass_LNC} and \ref{fig:Lambda_vs_mass_LNV}.
For all these results, we consider both LNC and LNV dimension-six operators, for all three quark-flavor scenarios.
In general, we find the constraints on the LNV operators to be 
slightly stronger than those on the LNC ones, because the meson decay branching ratios into HNLs in the LNV case are larger by about one order of magnitude than those in the LNC case (for the same value of the corresponding Wilson coefficients); 
see figures~\ref{fig:BrNNLNC} and \ref{fig:BrNNLNV}.

Among all the studied far detectors, we find MATHUSLA, ANUBIS, and AL3X show the best sensitivities, while FASER is usually the weakest.
Compared to FASER2, MAPP1 achieves very similar results with 100 times less data, mainly as a result of its much larger volume and closer distance to the IP.

In the $|V_{lN}|^2$ vs.~$m_N$ plots, we find these experiments can probe $|V_{lN}|^2$ as low as $10^{-15}$--$10^{-6}$, for GeV-scale HNLs.
Then, for two fixed benchmark values of $|V_{lN}|^2=10^{-7}$ and $|V_{lN}|^2 = 10^{-10}$, we obtain the sensitivity reach in the Wilson coefficients $c$ as functions of the HNL mass, for these far-detector experiments.
We further convert $c$ to the new-physics (NP) scale $\Lambda_\mathrm{NP}$.
We find for $|V_{lN}|^2=10^{-7}$, the considered experiments can probe the NP scale up to between approximately $10$ and several hundred TeV depending on the type of the operator considered, while for $|V_{lN}|^2=10^{-10}$, the reach in $\Lambda_\mathrm{NP}$ is reduced by slightly less than one order of magnitude, as a result of the larger decay length and hence smaller acceptance.

\section*{Acknowledgements}


This work is supported by the Spanish grants PID2020-113775GB-I00
(AEI/10.13039/ 501100011033) and CIPROM/2021/054 (Generalitat
Valenciana). R.B. acknowledges financial support from the Generalitat Valenciana (grant ACIF/2021/052).
G.C. acknowledges support from ANID FONDECYT grant No. 11220237.
G.C. and J.C.H. also acknowledge support from ANID
FONDECYT grant No. 1201673 and ANID -- Millennium Science Initiative
Program ICN2019\_044.  The work of A.T. is supported by the ``Generalitat
Valenciana'' under the grant PROMETEO/2019/087 
and by the MICINN-AEI (10.13039/501100011033) grant PID2020-113334GB-I00.
Z.S.W. is supported by the Ministry of Science and Technology (MoST) of Taiwan with grant number MoST-110-2811-M-007-542-MY3.

\appendix

\section{Pseudoscalar meson decays into HNLs}
\label{sec:decay_formulae}
The four-fermion interactions described by the $N_R$LEFT operators in tables~\ref{tab:opsNN} and \ref{tab:opsN} lead to two-body and three-body meson decays, 
if the HNL mass is small enough. 
The former involve two neutral leptons in the final state, whereas the latter correspond to semileptonic decays with an additional lighter meson within the final products. This appendix gathers the necessary formulae to determine meson decay rates into HNLs, assuming a pseudoscalar meson (of mass $m_P$ and momentum $p$) in the initial state.

In pure leptonic decays we define the meson decay constants through the transition matrix elements (see \textit{e.g.} Ref.~\cite{DeVries:2020jbs}):
\begin{align}
\langle 0 | \overline{q_i} \gamma^\mu \gamma_5 q_j | P(p) \rangle &= i f_P p^\mu\,, \nonumber \\
\langle 0 | \overline{q_i} \gamma_5 q_j | P(p) \rangle &= i \frac{m_P^2}{m_{q_i} + m_{q_j}} f_P \equiv  i f_P^S\,,
\end{align}
 where $i,j$ denote quark flavor indices, and $ |P(p) \rangle $ is the decaying pseudoscalar meson.
 
In semileptonic meson decays, 
the hadronic matrix elements are described by the corresponding form factors. 
We distinguish two scenarios depending on the nature of the outgoing meson. If it is a pseudoscalar (of mass $m$ and momentum $p'$), the non-vanishing matrix elements are:
\begin{align}
\langle P' (p') | \overline{q_i} \gamma^\mu q_j | P(p) \rangle &= f_{+}(q^2) \left[ (p + p')^{\mu} - \frac{m_P^2-m^2}{q^2} q^{\mu}\right] + f_{0}(q^2) \frac{m_P^2-m^2}{q^2} q^{\mu} \,, \nonumber  \\
\langle P' (p') | \overline{q_i} q_j | P(p) \rangle &= f_S(q^2) \,, \nonumber  \\
\langle P' (p') | \overline{q_i} \sigma^{\mu \nu} q_j | P(p) \rangle &= \frac{2i}{m_P+m} \left[ p^{\mu} p'^{\nu} - p^{\nu} p'^{\mu} \right] f_T(q^2) \,,
\label{eq:PPformfactors}
\end{align} 
where $q^{\mu} \equiv p^{\mu} - p'^{\mu}$. Of the four form factors ($f_+$, $f_0$, $f_S$ and $f_T$), the scalar one can be related to the others using the equations of motion, through
\begin{equation}
f_S(q^2) = \frac{m_P^2-m^2}{m_{q_j}-m_{q_i}} f_0(q^2) \,.
\end{equation}
When the final meson is a vector (of mass $m$, outgoing momentum $p'$ and polarization vector $\epsilon$), the non-zero matrix elements are:
\begin{align}
\langle V (p',\epsilon) | \overline{q_i} \gamma^{\mu} q_j | P(p) \rangle &= i g(q^2) \epsilon^{\mu \nu \alpha \beta} \epsilon_{\nu}^{\ast} P_{\alpha} q_{\beta} \,, \nonumber\\ 
\langle V (p',\epsilon) | \overline{q_i} \gamma^{\mu} \gamma_{5} q_j | P(p) \rangle &= f(q^2) \epsilon^{\ast \mu} + a_{+}(q^2) P^{\mu} \epsilon^{\ast} \cdot p + a_{-}(q^2) q^{\mu} \epsilon^{\ast} \cdot p \,, \nonumber \\
\langle V (p',\epsilon) | \overline{q_i} \sigma^{\mu \nu} q_j | P(p) \rangle &= \epsilon^{\mu \nu \alpha \beta}  \left( g_+(q^2) \epsilon_{\alpha}^{\ast} P_{\beta} + g_{-}(q^2) \epsilon_{\alpha}^{\ast} q_{\beta} + g_0(q^2) p_{\alpha} p'_{\beta} \, \epsilon^{\ast} \cdot p \right) \,, \nonumber  \\
\langle V (p',\epsilon) | \overline{q_i} \gamma_{5} q_j | P(p) \rangle &= f_{PS}(q^2)\, \epsilon^{\ast} \cdot p  \,,
\label{eq:PVformfactors}
\end{align}
where we defined $P^{\mu} \equiv p^{\mu} + p'^{\mu}$. These transitions, denoted by $P \rightarrow V$, are parametrized by eight functions of $q^2$: $g$, $f$, $a_+$, $a_-$, $g_+$, $g_-$, $g_0$ and $f_{PS}$. Equations of motion relate the pseudoscalar form factor with the rest through the following expression:
\begin{equation}
f_{PS} (q^2) = \frac{1}{m_{q_i}+m_{q_j}}
 \left[f(q^2) + a_+(q^2)\left(m_P^2-m^2\right) + a_-(q^2)q^2\right]. 
\end{equation}

\subsection{Two-body decays}
In the rest frame of a particle of mass $M$, the two-body decay width reads
\begin{equation}
 \Gamma =  \frac{\lambda^{1/2} \left(M^2,m_a^2,m_b^2\right)}{64\pi^2 M^3} \int \mathrm{d}\Omega_\mathrm{cm} \sum_{\mathrm{spins}}\left|\mathcal{M}\right|^2\,,
\end{equation}
where $\lambda \left(x,y,z \right) \equiv (x-y-z)^2 - 4yz$, and $m_a$, $m_b$ denote the masses of the decay products. Note that for identical particles in the final state one has $\int \mathrm{d}\Omega_{\mathrm{cm}} = 2\pi$.

In this section, we list the amplitudes and corresponding partial decay widths of a neutral pseudoscalar meson $P\sim \overline{q_i} q_j$ decaying into (i) a pair of HNLs, and (ii) one HNL plus one active neutrino. For a Dirac HNL (and Dirac $\nu$), these decays occur via the LNC pair-$N_R$ operators in table~\ref{tab:opsNN} and the LNC single-$N_R$ operators in table~\ref{tab:opsN}. In the Majorana scenario (where the HNL and $\nu$ are Majorana), the decays receive contributions from both the LNC and LNV operators. The amplitudes for Dirac neutrinos read
\begin{align}
 \mathcal{M}\left(P \to N\overline{N}\right) & = i \frac{f_P}{2} \left(c_{qN,\, ij}^{V,RR} - c_{qN,\,ij}^{V,LR}\right) \overline{u_N} \slashed{p} P_R v_{N'}\,, \\
  \mathcal{M}\left(P \to \nu_{\alpha} \overline{N}\right) & = i \frac{f_P^S}{2} \left(c_{q\nu N,\, ij\alpha}^{S,RR} - c_{q\nu N,\,ij\alpha}^{S,LR}\right) 
 \overline{u_\nu} P_R v_{N}\,, 
 \end{align}
where $u_N$ ($u_{\nu}$) and $v_{N}$ are the spinors corresponding to $N$ ($\nu$) and $\overline{N}$, respectively. A prime is added when there are two 
HNLs in the final state. For Majorana neutrinos, we recall that the full Majorana HNL field is $N= N_R^c + N_R$, such that $N^c = N$. 
The amplitude for $P \to N N$ has to be anti-symmetrized in $N$ and $N'$, since there are two identical particles in the final state.%
\footnote{The convention for associating a $u$ or $v$ spinor to a Majorana particle is arbitrary. Our is such that there is one $\overline{u}$ spinor and one $v$ spinor in each process. In this way, we recover the usual Dirac relations for the sums over spins. Nonetheless, the physical result does not depend on the convention's choice. }
 \begin{align}
 \mathcal{M}(P \to NN) &= i \frac{f_P}{2} \left(c_{qN,\, ij}^{V,RR} - c_{qN,\, ij}^{V,LR}\right) 
 \overline{u_N} \slashed{p} \gamma_5 v_{N'} \nonumber \\
 &+ i f_P^S \bigg[ \left(c_{qN,\,ij}^{S,RR} - c_{qN,\, ij}^{S,LR}\right) \overline{u_N} P_R v_{N'} 
 -  \left(c_{qN,\,ji}^{S,RR \, \ast} - c_{qN,\,ji}^{S,LR \, \ast}\right) \overline{u_N} P_L v_{N'} \bigg] \,, \\[0.2cm]
  \mathcal{M}\left(P \to \nu_{\alpha} N\right) &= i \frac{f_P^S}{2} \bigg[ \left(c_{q\nu N,\, ij\alpha}^{S,RR} - c_{q\nu N,\, ij\alpha}^{S,LR}\right)
 \overline{u_\nu} P_R v_N  - \left( c_{q\nu N,\, ji\alpha}^{S,RR \, \ast} - c_{q\nu N,\, ji\alpha}^{S,LR \, \ast} \right)  \overline{u_\nu} P_L v_N  \bigg]  \nonumber \\
 & + i \frac{f_P}{2} \bigg[ \left(c_{q\nu N,\, ij\alpha}^{V,RR}- c_{q\nu N,\, ij\alpha}^{V,LR}\right)
 \overline{u_\nu} \slashed{p} P_R v_N - \left(  c_{q\nu N,\, ji\alpha}^{V,RR \, \ast} - c_{q\nu N,\, ji\alpha}^{V,LR \, \ast}\right) \overline{u_\nu} \slashed{p} P_L v_N  \bigg]\,.
 \end{align}
The decay rates of these two processes in the Dirac scenario are given by
 \begin{align}
 \Gamma\left(P \to N\overline{N}\right) & = \frac{\left|f_P\right|^2}{32\pi} 
 \left|c_{qN,\, ij}^{V,RR} - c_{qN,\,ij}^{V,LR}\right|^2 m_P \, m_N^2 \, \sqrt{1-\frac{4m_N^2}{m_P^2}}\,, \\
 \Gamma\left(P \to \nu_\alpha \overline{N}\right)& =  \frac{\left|f_P^S\right|^2}{64 \pi} 
 \left|c_{q\nu N,\, ij\alpha}^{S,RR} - c_{q\nu N,\,ij\alpha}^{S,LR}\right|^2  m_P \left(1- \frac{m_N^2}{m_P^2}\right)^2\,, 
 \end{align}
and in the Majorana case by
 \begin{align}
 \Gamma\left(P \to NN\right) &= \frac{m_P}{32\pi} \sqrt{1-\frac{4m_N^2}{m_P^2}} 
 \Bigg[2 \left|f_P\right|^2 \left|c_{qN,\, ij}^{V,RR} - c_{qN,\,ij}^{V,LR}\right|^2 m_N^2 \nonumber \\ 
 &+ \left|f_P^S\right|^2\bigg\lbrace\left(\left|c_{qN,\, ij}^{S,RR} - c_{qN,\,ij}^{S,LR}\right|^2
 + \left|c_{qN,\, ji}^{S,RR} - c_{qN,\,ji}^{S,LR}\right|^2\right) \left(1 - \frac{2m_N^2}{m_P^2}\right) \nonumber \\
 &\hspace{1.5cm}+ 2\left[\left(c_{qN,\,ij}^{S,RR} - c_{qN,\, ij}^{S,LR}\right)\left(c_{qN,\,ji}^{S,RR} - c_{qN,\, ji}^{S,LR}\right) + \text{h.c.} \right] \frac{m_N^2}{m_P^2}\bigg\rbrace \nonumber \\
 &+ f_P f_P^{S} \bigg\lbrace  \left(c_{qN,\, ij}^{V,RR} - c_{qN,\,ij}^{V,LR}\right)
 \left(c_{qN,\,ij}^{S,RR \, \ast} - c_{qN,\,ij}^{S,LR \, \ast} + c_{qN,\,ji}^{S,RR} - c_{qN,\,ji}^{S,LR }\right) m_N 
 + \text{h.c.} \bigg\rbrace
 \Bigg]\,, 
 \label{eq:GammaPNNMaj}\\
 \Gamma\left(P \to \nu_\alpha N\right) &= \frac{m_P}{64\pi} \left(1-\frac{m_N^2}{m_P^2}\right)^2 
 \Bigg[ \left|f_P^S\right|^2 \left(\left|c_{q\nu N,\, ij\alpha}^{S,RR} - c_{q\nu N,\,ij\alpha}^{S,LR}\right|^2 + \left|c_{q\nu N,\, ji\alpha}^{S,RR} - c_{q\nu N,\,ji\alpha}^{S,LR}\right|^2 \right)\nonumber \\ 
 & + \left|f_P\right|^2 \left(\left|c_{q\nu N,\, ij\alpha}^{V,RR} - c_{q\nu N,\,ij\alpha}^{V,LR}\right|^2 +  \left|c_{q\nu N,\, ji\alpha}^{V,RR} - c_{q\nu N,\,ji\alpha}^{V,LR}\right|^2 \right) m_N^2 \nonumber \\
 & + f_P f_P^S  \bigg \lbrace \bigg[ \left(c_{q\nu N,\, ij\alpha}^{S,RR} - c_{q\nu N,\, ij\alpha}^{S,LR}\right) \left(c_{q\nu N,\, ji\alpha}^{V,RR}- c_{q\nu N,\, ji\alpha}^{V,LR}\right) \nonumber \\ 
 & \qquad \qquad +\left( c_{q\nu N,\, ji\alpha}^{S,RR \, \ast} - c_{q\nu N,\, ji\alpha}^{S,LR \, \ast} \right)  \left(  c_{q\nu N,\, ij\alpha}^{V,RR \, \ast} - c_{q\nu N,\, ij\alpha}^{V,LR \, \ast}\right) \bigg] m_N+ \text{h.c.} \bigg \rbrace\, \Bigg] \,.
 \label{eq:GammaPnuNMaj}
\end{align}

\subsection{Three-body decays}
\label{app:3-body}
For the three-body phase space integral calculations we follow the procedure described in Ref.~\cite[Appendix B.1]{DeVries:2020jbs}. However, we adopt general labels for the leptonic products. In the rest frame of the decaying pseudoscalar meson, the decay rate reads
\begin{align}
\Gamma & = \frac{1}{2m_P} \frac{1}{(2\pi)^5} \int \mathrm{d}^4 p' \int \mathrm{d}^4 p_a \int \mathrm{d}^4 p_b \, \delta(p'^2-m^2) \delta(p_a^2-m_a^2) \delta (p_b^2 - m_b^2) \nonumber \\[0.3em]
& \times \sum_{\mathrm{spins}}\left|\mathcal{M}\right|^2 \delta^{(4)}(p-p'-p_a-p_b) \,,
\end{align}
where $p$ ($p'$) is the momentum of the decaying (outgoing) meson and $p_a$ and $p_b$ denote the momenta of the produced leptons.%
\footnote{The labels $a$ and $b$ will refer to (i) $N'$ and $N$, and (ii) $\nu$ and $N$.} The summed over spins
squared matrix element can be written in terms of four-momentum invariant scalar products and the hadronic form factors, which are functions of $q^2 = (p-p')^2$.  After performing the integral over the four-momentum $p_b$, it is convenient to introduce the variable $a$ via a factor of $1= \int \mathrm{d}a \, \delta(a-q^2)$. Then, the spin-summed 
matrix element squared can be written as
\begin{equation}
\sum_\mathrm{spins} \left|\mathcal{M}\right|^2 \bigg \rvert_{p_b=q-p_a} = \sum_{n=0}^K c_n(a) (p_a \cdot p)^n \,,
\label{eq:matrixelement}
\end{equation}
where $c_n(a)$ are functions of $a$, particle masses and the hadronic form factors. For the decays under study we have $K \leq 2$. After performing the remaining momentum integrals, one arrives at the final decay rate expression\footnote{When the outgoing leptons are two Majorana HNLs an extra factor of $1/2$ appears in the final expression.}
\begin{equation}
\Gamma = \frac{1}{2m_P} \frac{1}{(2\pi)^5} \int^{(m_P-m)^2}_{(m_a+m_b)^2} \mathrm{d}a \, I_P \sum^K_{n=0} c_n(a) I_n \,,
\label{Gamma3body}
\end{equation}
where 
the integrals $I_P$ and $I_n$ (for $n={0,1,2}$) are given by
\begin{align}
I_P &= \frac{\pi}{2m_P^2} \lambda^{1/2}\left(a,m^2,m_P^2\right) \,, \nonumber \\
I_0 &= \frac{\pi}{2a} \lambda^{1/2}\left(a,m_a^2,m_b^2\right) \,, \qquad
I_1 = \frac{(p_a \cdot q)(p\cdot q)}{a} I_0 \,, \nonumber \\
I_2 &= \frac{I_0}{a^2} \left ( (p_a \cdot q)^2(p\cdot q)^2 + \frac{1}{48} \lambda(a,m^2,m_P^2) \lambda(a,m_a^2,m_b^2) \right) \,.
\end{align}
The scalar products above have to be replaced by 
\begin{equation}
p_a \cdot q = \frac{1}{2} \left(a + m_a^2 - m_b^2\right) \,, \qquad p\cdot q = \frac{1}{2} \left(m_P^2 + a - m^2 \right) \,.
\end{equation} 

In summary, after computing the spin-summed matrix element squared 
with standard techniques and arranging it in the form of Eq.\,\eqref{eq:matrixelement}, one replaces $(p_a \cdot p)^n$ by the corresponding integral result $I_n$, substitutes the form factor expressions (see section~\ref{sec:formfactors}), and evaluates the integral in Eq.~\eqref{Gamma3body} numerically.

In the rest of this section, we give the amplitudes of the relevant three-body decays for producing the HNLs via the $N_R$LEFT operators. For Dirac HNLs, these are

\begin{align}
\mathcal{M}\left(P \rightarrow P' N \overline{N} \right) & = \frac{1}{2}  \left(c_{qN,\, ij}^{V,RR} + c_{qN,\,ij}^{V,LR}\right) \left(f_{+} P_{\mu}  + \left(f_{0} - f_{+}\right)\frac{m_P^2-m^2}{q^2} q_{\mu}  \right)
 \left[ \overline{u_N} \gamma^{\mu} P_R v_{N'} \right]\,, \\
 \mathcal{M} \left(P \to V N\overline{N}\right)  &=  \bigg \lbrace \left(c_{qN,\, ij}^{V,RR} + c_{qN,\,ij}^{V,LR}\right) \left( i g \epsilon_{\mu \nu \alpha \beta} \epsilon^{\ast \nu} p'^{\alpha} p^{\beta} \right)  \nonumber \\
&+ \frac{1}{2} \left(c_{qN,\, ij}^{V,RR} - c_{qN,\,ij}^{V,LR}\right) \left( f \epsilon^{\ast}_{\mu} + a_{+} P_{\mu} \epsilon^\ast \cdot p + a_- q_{\mu} \epsilon^\ast \cdot p  \right) \bigg \rbrace
 \left[ \overline{u_N} \gamma^{\mu} P_R v_{N'} \right]\,, \\
 \mathcal{M} \left(P \to P' \nu_\alpha \overline{N}\right)  &=
\frac{f_S}{2} \left(c_{q\nu N,\, ij\alpha}^{S,RR} + c_{q\nu N,\, ij\alpha}^{S,LR}\right)  \left[\overline{u_\nu} P_R v_N \right] \nonumber \\
&+ c_{q\nu N,\,ij\alpha}^{T,RR} \frac{2i f_T}{m_P+m} \left(p_{\mu} p'_{\nu} + \frac{i}{2} \epsilon_{\mu\nu\rho\sigma} p^{\rho}p'^{\sigma} \right) \,\left[\overline{u_\nu} \sigma^{\mu \nu}P_R v_N \right]  \,, \\
\mathcal{M} \left(P \to V \nu_\alpha \overline{N}\right)  &=
\frac{f_{PS}}{2} \left(c_{q\nu N,\, ij\alpha}^{S,RR} - c_{q\nu N,\, ij\alpha}^{S,LR}\right)    (\epsilon^{\ast}\cdot p) \left[\overline{u_\nu} P_R v_N \right] \nonumber \\[3pt]
& + c_{q\nu N,\, ij\alpha}^{T,RR} \bigg \lbrace \frac{1}{2} \epsilon_{\mu \nu \alpha \beta} \left[ g_+ \epsilon^{\ast \alpha} (p + p')^\beta + g_- \epsilon^{\ast \alpha} q^\beta + g_0 p^\alpha p'^\beta  \epsilon^{\ast}\cdot p \right] \nonumber \\
& -i \left[ g_+ \epsilon^{\ast}_\mu (p + p')_\nu + g_- \epsilon^{\ast}_\mu q_\nu + g_0 p_\mu p'_\nu \epsilon^{\ast}\cdot p \right] \bigg \rbrace \left[\overline{u_\nu} \sigma^{\mu \nu}P_R v_N \right] \,.
\end{align}
The corresponding amplitudes for Majorana HNLs read
\begin{align}
\mathcal{M}(P\rightarrow P' NN) &= \frac{1}{2} \left(c_{qN,\, ij}^{V,RR} + c_{qN,\,ij}^{V,LR}\right) \left(f_{+} P_{\mu}  + \left(f_{0} - f_{+}\right)\frac{m_P^2-m^2}{q^2} q_{\mu} \right) \left[  \overline{u_N} \gamma^{\mu} \gamma^5 v_{N'}  \right] \nonumber \\
& + f_S \left(c_{qN,\, ij}^{S,RR} + c_{qN,\,ij}^{S,LR}\right) \left[  \overline{u_N} P_R v_{N'} \right] + f_S \left(c_{qN,\,ji}^{S,RR \ast} + c_{qN,\,ji}^{S,LR \ast}\right) \left[  \overline{u_N} P_L v_{N'} \right]  \,, 
\label{eq:MPPpNNMaj} \\
\mathcal{M}(P\rightarrow V NN)& = \bigg \lbrace \left(c_{qN,\, ij}^{V,RR} + c_{qN,\,ij}^{V,LR}\right) \left( i g \epsilon_{\mu \nu \alpha \beta} \epsilon^{\ast \nu} p'^{\alpha} p^{\beta} \right)  \nonumber \\
&+ \frac{1}{2} \left(c_{qN,\, ij}^{V,RR} - c_{qN,\,ij}^{V,LR}\right) \left( f \epsilon^{\ast}_{\mu} + a_{+} P_{\mu} \epsilon^\ast \cdot p + a_- q_{\mu} \epsilon^\ast \cdot p  \right) \bigg \rbrace
 \left[ \overline{u_N} \gamma^{\mu} \gamma^5 v_{N'} \right] \nonumber \\[3pt]
& + f_{PS} \bigg \lbrace \left(c_{q\nu N,\, ij}^{S,RR}- c_{q\nu N,\, ij}^{S,LR}\right) 
(\epsilon^{\ast} \cdot p) \left[ \overline{u_N} P_R v_{N'}\right] \nonumber \\
&\hspace{1.5cm} - \left( c_{q\nu N,\, ji}^{S,RR \, \ast} - c_{q\nu N,\, ji}^{S,LR \, \ast} \right) 
(\epsilon^{\ast} \cdot p) \left[ \overline{u_N} P_L v_{N'}\right]  \bigg \rbrace \,, 
\label{eq:MPVNNMaj} \\
 \mathcal{M}(P\rightarrow P' \nu_\alpha N) & =  \frac{f_S}{2} \bigg \lbrace \left(c_{q\nu N,\,ij\alpha}^{S,RR} + c_{q\nu N,\,ij\alpha}^{S,LR}\right) \left[\overline{u_\nu} P_R v_N \right] +  \left(c_{q\nu N,\,ji\alpha}^{S,RR\, \ast} + c_{q\nu N,\,ji\alpha}^{S,LR \, \ast}\right) \left[\overline{u_\nu} P_L v_N \right] \bigg \rbrace \nonumber \\[5pt]
& + \frac{i f_T}{M+m} \bigg \lbrace c_{q\nu N,\,ij\alpha}^{T,RR} \left(2 p_{\mu} p'_{\nu} + i \epsilon_{\mu \nu \rho \sigma} p^{\rho} p'^{\sigma} \right) \left[\overline{u_\nu} \sigma^{\mu \nu}P_R v_N \right] \nonumber \\[5pt]
&\hspace{2cm} - c_{q\nu N,\,ji\alpha}^{T,RR\, \ast} \left(2 p_{\mu} p'_{\nu} - i \epsilon_{\mu \nu \rho \sigma} p^{\rho} p'^{\sigma} \right) \left[\overline{u_\nu} \sigma^{\mu \nu}P_L v_N \right]\bigg \rbrace \nonumber \\[5pt]
& + \bigg \lbrace \left(c_{q\nu N,\,ij\alpha}^{V,RR}+ c_{q\nu N,\,ij\alpha}^{V,LR}\right)  \left[ \overline{u_\nu} \gamma^{\mu} P_R v_N \right] 
  - \left(c_{q\nu N,\,ji\alpha}^{V,RR \, \ast}+ c_{q\nu N,\,ji\alpha}^{V,LR \, \ast}\right) \left[ \overline{u_\nu} \gamma^{\mu} P_L v_N \right] \bigg \rbrace \nonumber \\[5pt]
&\times \frac{1}{2} \left(f_{+} P_{\mu}  + \left(f_{0} - f_{+}\right)\frac{m_P^2-m^2}{q^2} q_{\mu}  \right)\,, \\[1em]
 \mathcal{M}\left( P \rightarrow V \nu_\alpha N \right) & = \frac{f_{PS}}{2} 
\bigg \lbrace \left(c_{q\nu N,\,ij\alpha}^{S,RR}-c_{q\nu N,\,ij\alpha}^{S,LR}\right) 
(\epsilon^\ast \cdot p) \left[\overline{u_\nu} P_R v_N \right] \nonumber \\
&\hspace{1.5cm} - \left(c_{q\nu N,\,ji\alpha}^{S,RR \,\ast}-c_ {q\nu N,\,ji\alpha}^{S,LR \, \ast}\right) 
(\epsilon^\ast \cdot p) \left[\overline{u_\nu} P_L v_N \right] \bigg \rbrace  \nonumber \\
&+ c_{q\nu N,\,ij\alpha}^{T,RR} \left( \frac{1}{2}\epsilon_{\mu \nu \alpha \beta} T^{\alpha \beta} - i T_{\mu \nu}\right) \left[\overline{u_\nu} \sigma^{\mu \nu} P_R v_N \right] \nonumber \\
&- c_{q\nu N,\,ji \alpha}^{T,RR \, \ast} \left( \frac{1}{2}\epsilon_{\mu \nu \alpha \beta} T^{\alpha \beta} + i T_{\mu \nu}\right) \left[\overline{u_\nu} \sigma^{\mu \nu} P_L v_N \right]    \nonumber \\
& + \frac{1}{2} \bigg\lbrace \left(c_{q\nu N,\,ij\alpha}^{V,RR}+ c_{q\nu N,\,ij\alpha}^{V,LR}\right) V_{\mu} + \left(c_{q\nu N,\,ij\alpha}^{V,RR}- c_{q\nu N,\,ij\alpha}^{V,LR}\right) A_{\mu} \bigg\rbrace \left[ \overline{u_\nu} \gamma^{\mu} P_R v_N\right] \nonumber \\
&  -\frac{1}{2} \bigg\lbrace \left(c_{q\nu N,\,ji\alpha}^{V,RR\, \ast}+ c_{q\nu N,\,ji\alpha}^{V,LR\, \ast}\right) V_{\mu}  + \left(c_{q\nu N,\,ji\alpha}^{V,RR\, \ast}- c_{q\nu N,\,ji\alpha}^{V,LR \, \ast}\right) A_{\mu}  \bigg\rbrace \left[ \overline{u_\nu} \gamma^{\mu} P_L v_N \right]\,,
\end{align}
where we have introduced:
\begin{align*}
T^{\alpha \beta} & \equiv g_+ \epsilon^{\ast \, \alpha} (p + p')^\beta + g_- \epsilon^{\ast \, \alpha} q^\beta + g_0 p^\alpha p'^\beta (\epsilon^{\ast} \cdot p) \,, \\
V_{\mu} &\equiv 2 i g \epsilon_{\mu \nu \alpha \beta} \epsilon^{\ast \nu}  p'^{\alpha}p^\beta \,, \\
A_\mu &\equiv f \epsilon^{\ast}_{\mu} + a_+ (p+p')_{\mu}(\epsilon^\ast \cdot p) + a_- q_\mu (\epsilon^\ast \cdot p )\,.
\end{align*}
Specific software, such as \verb|FeynCalc|~\cite{Shtabovenko:2016sxi,Shtabovenko:2020gxv}, automatizes the computation of the summed over spins squared matrix elements. Finally, to simplify notation in the previous expressions, we have omitted the $q^2$ dependence of the form factors.

\subsection{Decay constants and form factors}
\label{sec:formfactors}
The masses of all mesons involved in the decays are taken from Ref.~\cite{Workman:2022ynf}. 
In addition, we use the light quark masses at a renormalization scale of $\mu = 2$~GeV in $\overline{\text{MS}}$, 
whereas for $m_c$ and $m_b$ we employ the $\overline{\text{MS}}$ masses
at $\mu = m_c$ and $\mu = m_b$, respectively,
as in Ref.~\cite{Workman:2022ynf}:
\begin{equation*}
 m_u = 2.2 \text{ MeV}, \quad m_d=4.7 \text{ MeV}, \quad m_s = 93 \text{ MeV} , \quad
m_c = 1.27 \text{ GeV}, \quad m_b = 4.18 \text{ GeV}.
\end{equation*}

We provide the decay constants of the heavy 
neutral pseudoscalar mesons in table~\ref{tab:decayconstants}. 
We use the isospin-averaged results%
\footnote{In other words, we assume the isospin symmetry, 
\textit{i.e.} the (approximate) symmetry between the $u$ and $d$ quarks. 
In this limit, the decay constants for $D^0$ and $D^+$ are the same, $f_D$. 
Similarly, $B^0$ and $B^+$ mesons are characterised by one constant, $f_B$. The difference $|f_{P^+} - f_P|$ is estimated to be $\simeq 0.5$~MeV for both $P=D$ and $P=B$~\cite{Aoki:2021kgd,PDG2021update}.} 
for $N_f = 2+1+1$ dynamical quark flavors from Ref.~\cite{Aoki:2021kgd}.
\begin{table}[t]
 \centering
 \begin{tabular}{| c | c |}
 \hline
 Meson $P$  & Decay constant $f_P$~[MeV] \\
 \hline
 $D^0$ &  212.0 (0.7) \\
 $B^0$ &  190.0 (1.3) \\
 $B^0_s$ & 230.3 (1.3) \\
 \hline
 \end{tabular}
 \caption{Decay constants of heavy neutral pseudoscalar mesons \cite{Aoki:2021kgd}.}
 \label{tab:decayconstants}
\end{table}
As regards hadronic form factors, several parametrizations exist in the literature for describing their dependence on $q^2$. One of the most used is the BCL method \cite{Bourrely:2008za}, and we try to stick to it as long as there is available data. In all other cases, we use the so-called double-pole parametrization. In table~\ref{tab:formfactors}, we show all hadronic transitions mediated by the selected $N_R$LEFT operators and the references studying the associated form factors. We indicate the equation where the explicit $q^2$ parametrization appears and the tables where the best-fit parameter values are given.%
\footnote{Notice that there are several conventions in the literature for writing hadronic matrix elements in terms of form factors. The basis presented in Eqs.~\eqref{eq:PPformfactors} and \eqref{eq:PVformfactors} must be converted to the new basis in each reference before using their $q^2$ parametrization.}
%
\begin{table}[t]
\centering
\begin{tabular}{|ll|c|c|}
\hline
\multicolumn{2}{|c|}{Meson transitions }                     &        Form factor parametrization            &  Fitted parameters                 \\ \hline
\multicolumn{1}{|l|}{\multirow{5}{*}{$D$-mesons}} & $D^0 \rightarrow \pi^0$ & \multirow{2}{*}{BCL \cite[Eqs.~(A1)--(A2)]{Lubicz:2017syv}} & \multirow{2}{*}{\cite[Tab.~VI]{Lubicz:2017syv}} \\ 
\multicolumn{1}{|l|}{} & $D^+ \rightarrow \pi^+$ &                      &                   \\ \cline{2-4} 
\multicolumn{1}{|l|}{}                  & $D^0 \rightarrow \eta,\,\eta',\,\rho^0,\,\omega$ & \multirow{3}{*}{Double-pole \cite[Eq.~(63)]{Ivanov:2019nqd}} & \multirow{3}{*}{\cite[Tab.~VI]{Ivanov:2019nqd}} \\
\multicolumn{1}{|l|}{}                  & $D^+ \rightarrow \rho^+$ &                  &                   \\
\multicolumn{1}{|l|}{}                  & $D_s^+ \rightarrow K^+,\,K^{\ast +} $ &                     &                   \\ \hline
\multicolumn{1}{|l|}{\multirow{8}{*}{$B$-mesons}} & $B^0 \rightarrow \pi^0,\,K^0$ & \multirow{3}{*}{BCL \cite[Eqs.~(529)--(530)]{Aoki:2021kgd}} & \multirow{3}{*}{\cite[Tabs.~46,\,48,\,50,\,51]{Aoki:2021kgd}} \\
\multicolumn{1}{|l|}{}                  &  $B^+ \rightarrow \pi^+,\,K^+$&                             &                   \\
\multicolumn{1}{|l|}{}                  & $B_s^0 \rightarrow \overline{K^0}$ &                         &                   \\ \cline{2-4} 
\multicolumn{1}{|l|}{}                  & $B^0 \rightarrow \eta,\,\eta'$  & \multirow{2}{*}{Double-pole \cite[Eq.~(36)]{Wu:2006rd}} & \multirow{2}{*}{\cite[Tab.~4]{Wu:2006rd}} \\
\multicolumn{1}{|l|}{}                  & $B_s^0 \rightarrow \eta,\,\eta'$ &                     &                   \\ \cline{2-4} 
\multicolumn{1}{|l|}{}                  & $B^0 \rightarrow \rho^0,\,\omega,\,K^{\ast 0}$   & \multirow{3}{*}{BCL \cite[Eq.~(2.16)]{Bharucha:2015bzk}} & \multirow{3}{*}{\cite[Tab.~14]{Bharucha:2015bzk}} \\
\multicolumn{1}{|l|}{}                  & $B^+ \rightarrow  \rho^+,\,K^{ \ast +}$ &                             &                   \\
\multicolumn{1}{|l|}{}                  &  $B_s^0 \rightarrow \phi,\,\overline{K^{\ast 0}}$ &                        &                   \\ \hline
\end{tabular}
 \caption{Relevant meson transitions for the decays studied in section~\ref{sec:hnlproduction}. The two last columns show the references to the form factor parametrizations and the best-fit parameter values. Isospin symmetry is assumed so that identical form factors are taken for $P^0 \rightarrow M^0$ and $P^+ \rightarrow M^+$ transitions, where $M$ can be a pseudoscalar or vector meson. }
 \label{tab:formfactors}
\end{table}
%

\section{Branching ratios of meson decays triggered by single-$N_R$ operators}
\label{sec:BRnuN}
\begin{figure}[t]
 %
 %
 \includegraphics[height=0.27\textheight]{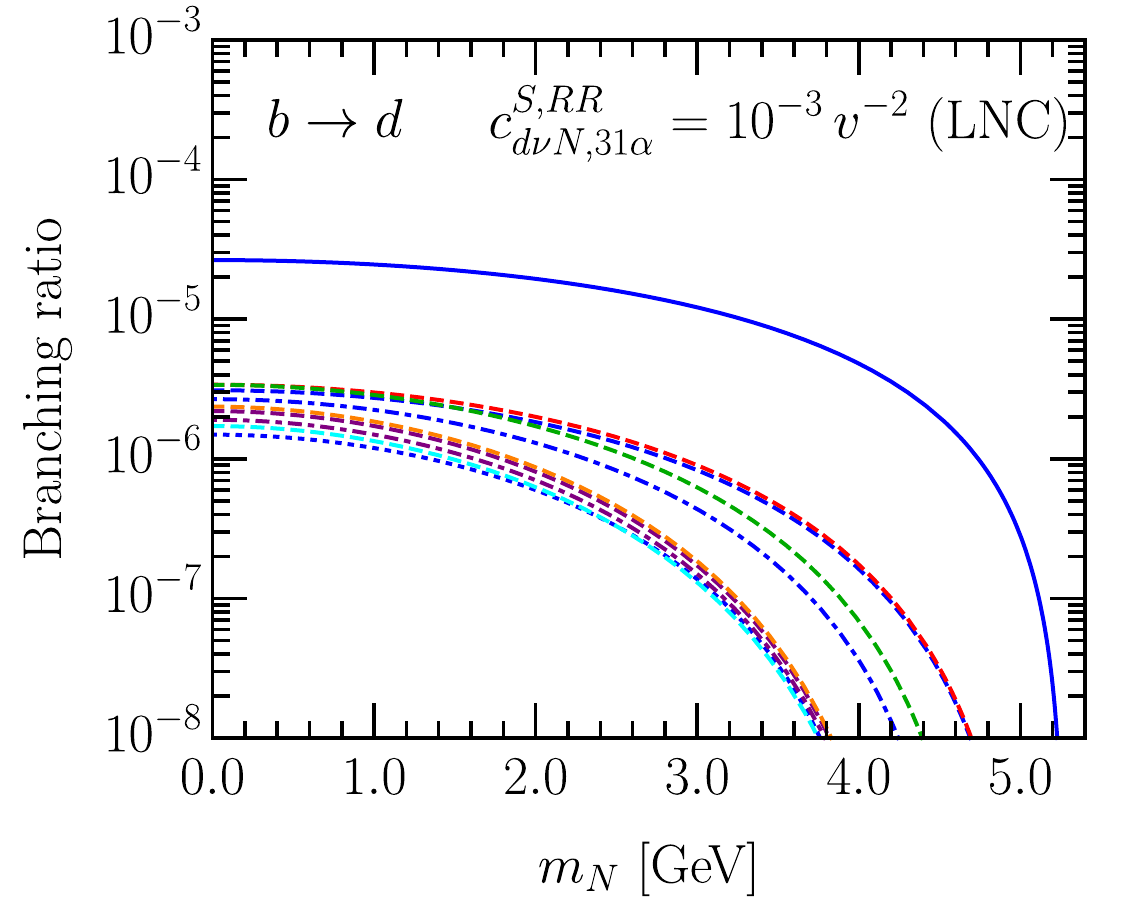}
 \includegraphics[height=0.27\textheight]{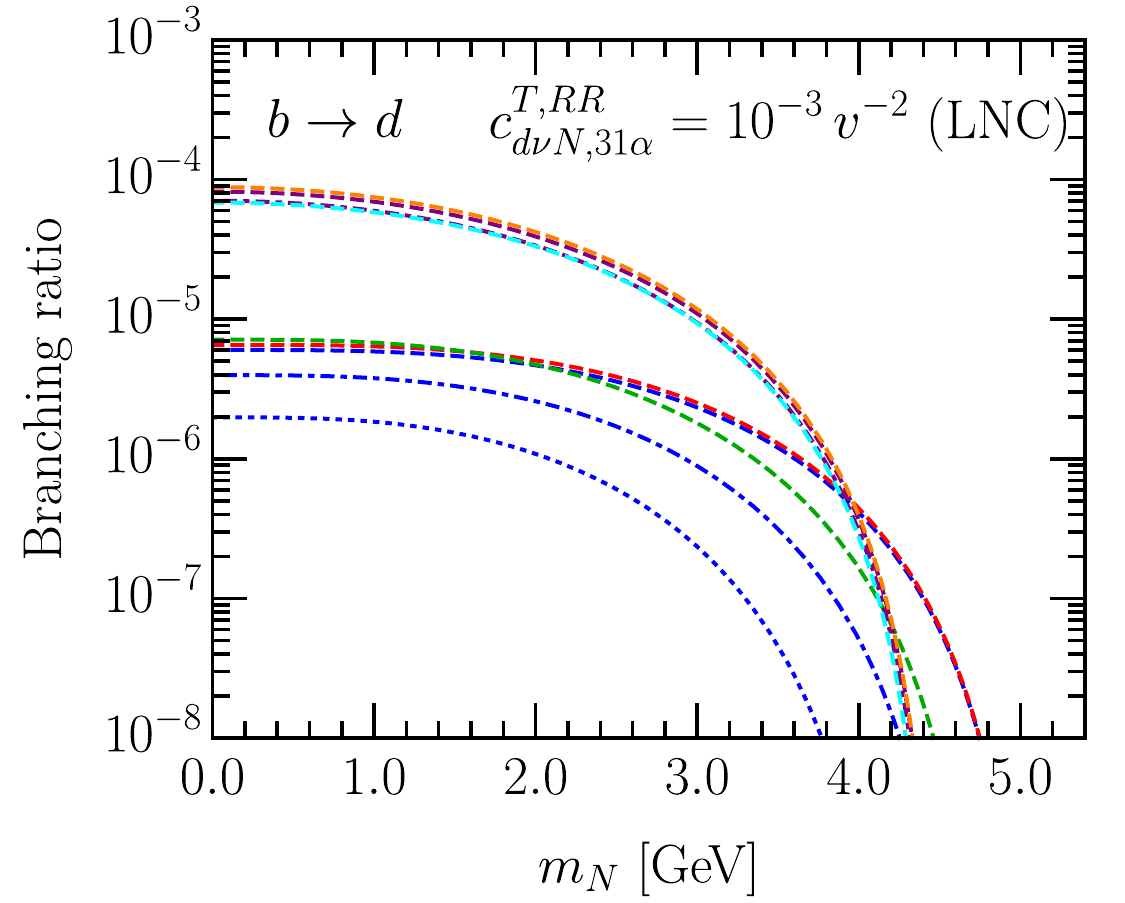}
 \includegraphics[height=0.27\textheight]{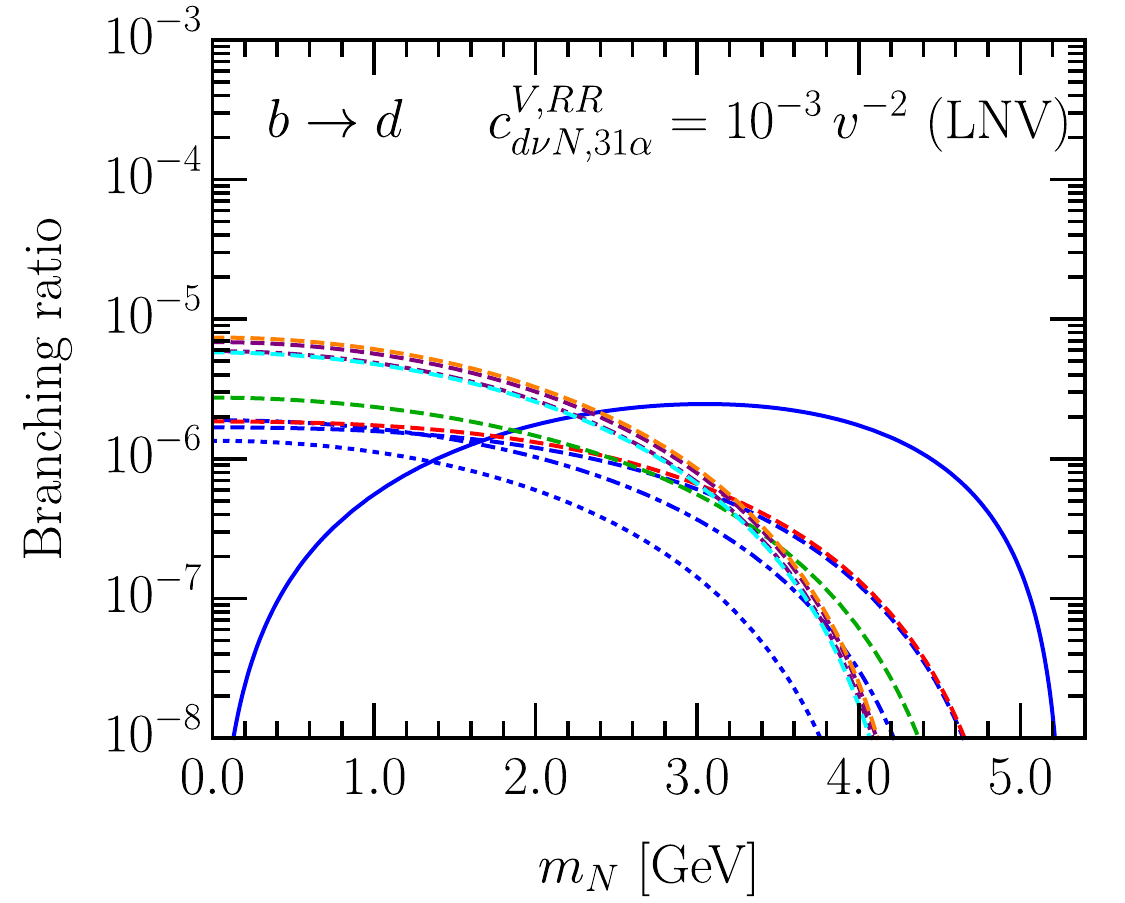}
 \hspace{1cm}
 \includegraphics[height=0.27\textheight]{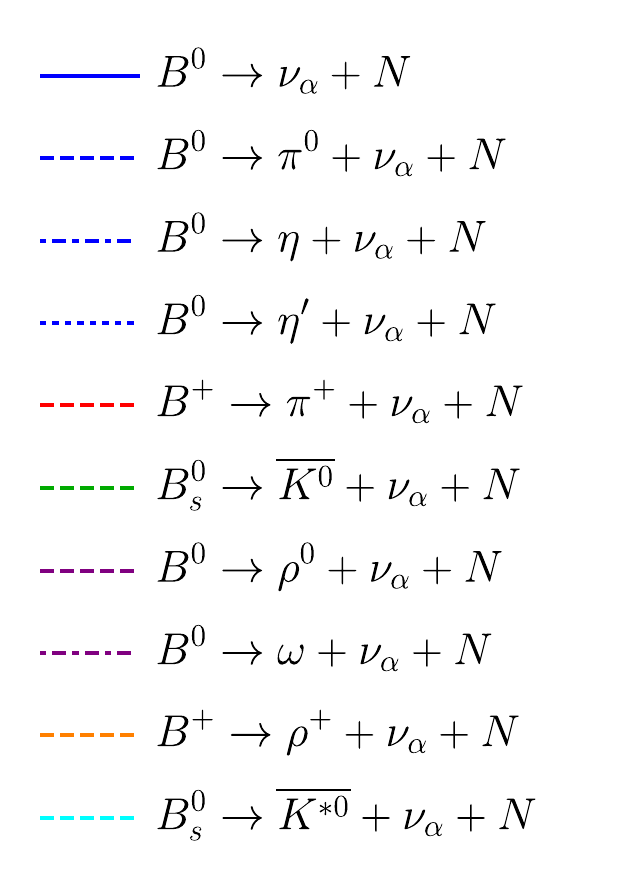}
 \caption{Branching ratios of $B$-meson decays triggered 
by the single-$N_R$ operators $\mathcal{O}_{d\nu N,31\alpha}^{S,RR}$, 
$\mathcal{O}_{d\nu N,31\alpha}^{T,RR}$
and $\mathcal{O}_{d\nu N,32\alpha}^{V,RR}$. 
In each case, the corresponding Wilson coefficient has been set to $10^{-3} v^{-2}$.}
 \label{fig:BrBnuN}
\end{figure}

%
We have also computed the branching ratios for 
two- and three-body pseudoscalar meson decays 
including in the final state a SM neutrino and an HNL.
These decays are triggered by the single-$N_R$ operators given in table~\ref{tab:opsN}. 
As an example, we show in figure~\ref{fig:BrBnuN} 
the corresponding branching ratios in the case of $b \to d$ transitions 
triggered by three single-$N_R$ operators, namely, 
$\mathcal{O}_{d\nu N,31\alpha}^{S,RR}$, 
$\mathcal{O}_{d\nu N,31\alpha}^{T,RR}$, and 
$\mathcal{O}_{d\nu N,31\alpha}^{V,RR}$. 
The scalar and tensor operators are LNC, whereas the operator 
of the vector type is LNV. 
%

As can be seen from Eq.~\eqref{eq:GammaPnuNMaj}, the tensor operator 
does not contribute to the two-body decay width. This is why there is no solid 
blue line in the corresponding plot in figure~\ref{fig:BrBnuN}.
The tensor form factors for the $B \to \pi$ transitions have been taken 
from table~50 in Ref.~\cite{Aoki:2021kgd}. For $B^0 \to \eta,~\eta'$ and $B_s^0 \to \overline{K^0}$, we have used 
Eq.~(27) from Ref.~\cite{Wu:2006rd}, which relates $f_T$ to $f_+$ and $f_0$. 
Finally, for the $B_{(s)} \to V$ transitions, we have used the results of Ref.~\cite{Bharucha:2015bzk}. In table~\ref{tab:formfactors} we include these references and the ones for all other form factors.

We note that the branching ratios in figure~\ref{fig:BrBnuN} are very similar 
to those for the meson decays induced by the single-$N_R$ operators
with a charged lepton, which have been studied in detail in Ref.~\cite{DeVries:2020jbs}, 
cf.~figure~2 therein. Thus, the constraints on the neutral current 
single-$N_R$ operators from the far detectors
are expected to be very similar to those derived in Ref.~\cite{DeVries:2020jbs} 
for the charged current single-$N_R$ operators. 
For this reason, we do not study the neutral current single-$N_R$ operators in more detail.

\section{HNL two-body decays via single-$N_R$ operators}
\label{sec:decaysingleN}
Single-$N_R$ operators contribute to both HNL production and decay. In fact, depending on the Wilson coefficient and the value of 
active-heavy mixing, they can even dominate the HNL decay width. The LNC and LNV operators in table~\ref{tab:opsN} make HNLs undergo two- and three-body semileptonic decays. The dominant channels are two-body decays, mediated by just one insertion of an effective operator, with one light neutral meson and one active neutrino in the final state. 
If the HNL mass is large enough, then multimeson final states could also be important in the computation of the total decay width. We do not compute this contribution, though.
The single-$N_R$ operators with light quarks can mediate the two-body decay processes mentioned above at tree level. The contribution from operators with heavy quarks, which control $B$- and $D$-meson decays into HNLs, is loop-suppressed, 
and we do not compute it. The interference between active-heavy mixing and single-$N_R$ operators is also not treated here.

In this appendix, we list the HNL partial decay rates into a neutral meson and 
a SM neutrino mediated by the effective operators containing two light quarks. 
We leave, however,  the flavor indices of the Wilson coefficients unspecified. We consider the scenarios where both $N$ and $\nu$ are either Dirac or Majorana particles. In each case, we distinguish between a pseudoscalar and a vector meson in the final state. The latter requires the introduction of the following form factors:
\begin{align}
\langle 0 | \overline{q_i} \gamma^\mu q_j | V(p,\epsilon) \rangle & = i f_V m_V \epsilon^{\mu} \,, \nonumber \\
\langle 0 | \overline{q_i} \sigma ^{\mu \nu} q_j | V(p,\epsilon) \rangle & = - f_V^T (p^\mu \epsilon^\nu - p^\nu \epsilon^\mu)\,,
\end{align}
where $|V(p,\epsilon)\rangle$ denotes a vector meson with mass $m_V$, momentum $p$, and polarization vector $\epsilon$. For Dirac $N$ and $\nu$, the decay rate into 
a SM neutrino and a pseudoscalar (vector) meson $P~(V)\sim \overline{q_i} q_j$ of mass $m_P$ ($m_V$) reads
\begin{align}
\Gamma (N \rightarrow \nu_\alpha P) & = \frac{\left|f_P^S\right|^2}{128 \pi} \left|c_{q\nu N,\,ji\alpha}^{S,RR} - c_{q\nu N,\,ji\alpha}^{S,LR} \right|^2 m_N \left(1-\frac{m_P^2}{m_N^2}\right)^2 \,, \\
\Gamma (N \rightarrow \nu_\alpha V) & = \frac{\left|f_V^T\right|^2}{4 \pi} \left|c_{q\nu N,\,ji\alpha}^{T,RR}\right|^2 m_N^3 \left( 1 - \frac{m_V^2}{m_N^2}\right)^2 \left(1+ \frac{m_V^2}{2m_N^2}\right) \,.
\end{align}
In the Majorana scenario, the decay rates receive contributions from both the LNC and LNV single-$N_R$ operators and are given by
\begin{align}
\Gamma (N \rightarrow \nu_\alpha P) & = \frac{m_N}{128\pi} \left(1-\frac{m_P^2}{m_N^2}\right)^2 \Bigg \lbrace \left|f_P^S\right|^2   \left( \left|c_{q\nu N,\,ji\alpha}^{S,RR} - c_{q\nu N,\,ji\alpha}^{S,LR}\right|^2 + \left|c_{q\nu N,\,ij\alpha}^{S,RR} - c_{q\nu N,\,ij\alpha}^{S,LR}\right|^2 \right)  \nonumber \\
& +  \left|f_P\right|^2 \left( \left|c_{q\nu N,\,ji\alpha}^{V,RR} - c_{q\nu N,\,ji\alpha}^{V,LR}\right|^2 + \left|c_{q\nu N,\,ij\alpha}^{V,RR} - c_{q\nu N,\,ij\alpha}^{V,LR}\right|^2 \right) m_N^2  \nonumber \\
& -  f_P f_P^S \bigg[ \left(c_{q\nu N,\,ji\alpha}^{S,RR} - c_{q\nu N,\,ji\alpha}^{S,LR}\right)\left(c_{q\nu N,\,ij\alpha}^{V,RR} - c_{q\nu N,\,ij\alpha}^{V,LR}\right)  \nonumber \\
& \hspace{1.4cm} + \left(c_{q\nu N,\,ij\alpha}^{S,RR} - c_{q\nu N,\,ij\alpha}^{S,LR}\right)\left(c_{q\nu N,\,ji\alpha}^{V,RR} - c_{q\nu N,\,ji\alpha}^{V,LR}\right) + \text{h.c.} \bigg] m_N \Bigg \rbrace \,, \\
\Gamma (N \rightarrow \nu_\alpha V) & = \frac{m_N^3}{128\pi} \left(1- \frac{m_V^2}{m_N^2}\right)^2 \Bigg \lbrace 32 \left|f_V^T\right|^2 
\left(\left|c_{q\nu N,\,ij\alpha}^{T,RR}\right|^2 + \left|c_{q\nu N,\,ji\alpha}^{T,RR}\right|^2\right) \left(1+ \frac{m_V^2}{2m_N^2}\right) \nonumber \\
& + \left|f_V\right|^2 \left( \left|c_{q\nu N,\,ji\alpha}^{V,RR} + c_{q\nu N,\,ji\alpha}^{V,LR}\right|^2 + \left|c_{q\nu N,\,ij\alpha}^{V,RR} + c_{q\nu N,\,ij\alpha}^{V,LR}\right|^2 \right) \left(1+ \frac{2m_V^2}{m_N^2} \right) \nonumber \\
& - 12 f_V^T f_V \bigg[c_{q\nu N,\,ji\alpha}^{T,RR} \left(c_{q\nu N,\,ij\alpha}^{V,RR} + c_{q\nu N,\,ij\alpha}^{V,LR}\right) \nonumber \\
& \hspace{1.8cm} + c_{q\nu N,\,ij\alpha}^{T,RR} \left(c_{q\nu N,\,ji\alpha}^{V,RR} + c_{q\nu N,\,ji\alpha}^{V,LR}\right) + \text{h.c.} \bigg]\frac{m_V}{m_N} \Bigg \rbrace\,.
\end{align}

\bibliographystyle{JHEP}
\bibliography{RefsEFT}

\end{document}